%% file: main.tex
\documentclass[sigconf]{acmart}

\copyrightyear{2021}
\acmYear{2021}
\setcopyright{acmcopyright}\acmConference[ESEC/FSE '21]{Proceedings of the 29th
ACM Joint European Software Engineering Conference and Symposium on the
Foundations of Software Engineering}{August 23--27, 2021}{Athens, Greece}
\acmBooktitle{Proceedings of the 29th ACM Joint European Software Engineering
Conference and Symposium on the Foundations of Software Engineering (ESEC/FSE
'21), August 23--27, 2021, Athens, Greece}
\acmPrice{15.00}
\acmDOI{10.1145/3468264.3468555}
\acmISBN{978-1-4503-8562-6/21/08}

\usepackage{pgfplots}
\usepackage{prettyref}
\usepackage{tikz}
\usepackage{xcolor,colortbl}
\usepackage{multirow}
\usepackage[ruled, linesnumbered, vlined, commentsnumbered]{algorithm2e}
\usepackage[section]{placeins}
\usepackage{booktabs}
\usepackage{threeparttable}
\usepackage{makecell}
\usepackage{graphicx}
\usepackage[listings,skins,breakable]{tcolorbox}
\usepackage{csquotes}
\usepackage{subcaption}
\usepackage{standalone}
\usepackage[skip=0.5\baselineskip]{caption}
\usepackage{wrapfig}
\usepackage{amsmath}
\usetikzlibrary{pgfplots.statistics}
\usepackage{tcolorbox}
\usepackage{pifont}
\usepackage{balance}
\newtcolorbox{quotebox}{colback=steel!10,boxrule=0.4pt,colframe=black,fonttitle=\bfseries,top=2pt,bottom=2pt}

\mathchardef\mhyphen="2D
\newcommand{\vect}[1]{\boldsymbol{#1}}
\DeclareMathAlphabet\mathbfcal{OMS}{cmsy}{b}{n}

\newbox\aMark
\setbox\aMark\hbox{\begin{pgfpicture}\textcolor{red}{\pgfuseplotmark{o}}\end{pgfpicture}}

\newbox\bMark
\setbox\bMark\hbox{\begin{pgfpicture}\textcolor{red}{\pgfuseplotmark{star}}\end{pgfpicture}}

\newrefformat{fig}{Fig.~\ref{#1}}
\newrefformat{tab}{Table~\ref{#1}}
\newrefformat{sec}{Section~\ref{#1}}
\newrefformat{app}{Appendix~\ref{#1}}
\newrefformat{alg}{Algorithm~\ref{#1}}
\newrefformat{property}{Property~\ref{#1}}
\newrefformat{theorem}{Theorem~\ref{#1}}
\newrefformat{lemma}{Lemma~\ref{#1}}
\newrefformat{corollary}{Corollary~\ref{#1}}
\newrefformat{proposition}{Proposition~\ref{#1}}
\newrefformat{def}{Definition~\ref{#1}}
\newrefformat{eq}{equation~(\ref{#1})}

\definecolor{steel}{rgb}{0, 0.2, 0.9} 

\pgfplotsset{compat=newest}

\pgfplotsset{compat=1.11,
    /pgfplots/ybar legend/.style={
    /pgfplots/legend image code/.code={%
       \draw[##1,/tikz/.cd,yshift=-0.25em]
        (0cm,0cm) rectangle (3pt,0.8em);},
   },
}

\DeclareMathOperator*{\argmin}{argmin}

\def\signed #1{{\leavevmode\unskip\nobreak\hfil\penalty50\hskip2em
  \hbox{}\nobreak\hfil(#1)%
  \parfillskip=0pt \finalhyphendemerits=0 \endgraf}}

\newsavebox\mybox
\newenvironment{aquote}[1]
  {\savebox\mybox{#1}\begin{quote}}
  {\signed{\usebox\mybox}\end{quote}}

\begin{document}
	



\title[Multi-Objectivizing Software Configuration Tuning]{Multi-Objectivizing Software Configuration Tuning}

\subtitle{for a Single Performance Concern}


\author{Tao Chen}
\authornote{Both authors contributed equally to this research.}
\affiliation{%
  \institution{Loughborough University}
  \state{Loughborough}
  \country{United Kingdom}}
  \email{t.t.chen@lboro.ac.uk}

  \author{Miqing Li}
\affiliation{%
  \institution{University of Birmingham}
  \state{Birmingham}
  \country{United Kingdom}}
  \email{m.li.8@cs.bham.ac.uk}



\begin{abstract}

Automatically tuning software configuration for optimizing a single performance attribute
(e.g., minimizing latency) 
is not trivial, 
due to the nature of the configuration systems (e.g., complex landscape and expensive measurement). To deal with the problem, existing work has been focusing on developing various effective optimizers.
However, a prominent issue that all these optimizers need to take care of is how to avoid the search being trapped in local optima --- 
a hard nut to crack for software configuration tuning 
due to its rugged and sparse landscape, 
and neighboring configurations tending to behave very differently. 
Overcoming such in an expensive measurement setting is even more challenging.
In this paper, 
we take a different perspective to tackle this issue. 
Instead of focusing on improving the optimizer, 
we work on the level of optimization model. 
We do this by proposing a meta multi-objectivization model (MMO) 
that considers an auxiliary performance objective (e.g., throughput in addition to latency).
What makes this model unique is that we do not optimize the auxiliary performance objective, 
but rather use it to make similarly-performing while different configurations less comparable 
(i.e. Pareto nondominated to each other), 
thus preventing the search from being trapped in local optima.


Experiments on eight real-world software systems/environments with diverse performance attributes reveal that 
our MMO model is statistically more effective than state-of-the-art single-objective counterparts 
in overcoming local optima (up to 42\% gain), 
while using as low as 24\% of their measurements to achieve the same (or better) performance result. 

\end{abstract}

\maketitle

\input{introduction}

\input{problem}

\input{method}

\input{study}

\input{results}

\input{discussion}

\input{related}

\input{conclusion}

\begin{acks}
The authors would like to thank the reviewers for their constructive and insightful comments on helping improve the work.
\end{acks}

\balance
\bibliographystyle{ACM-Reference-Format}
\bibliography{reference} 
\citestyle{acmauthoryear}

\end{document}

%% file: introduction.tex
\section{Introduction}
\label{sec:intro}

\begin{aquote}{\textit{An anonymous industry partner}}
\textit{``All I want is to optimize the latency of my software; any other performance attributes are out of interest.''}
\end{aquote}
The above quotation comes from one of our industry partners who is working in the finance sector, 
commenting on the need of tuning the configuration of a software system that manages all financial trading in his company. 
In this case, only a single performance attribute matter (i.e., latency) --- 
in the finance sector, 
a millisecond decrease in the trade delay may boost a high-speed firm's earnings by about 100 million USD per year~\cite{tian2015latency}.

Indeed, given the flexibility of highly-configurable software systems, automatically tuning their critical configuration options will affect a set of performance attributes, such as latency, throughput, and energy consumption~\cite{DBLP:conf/sigsoft/ShahbazianKBM20,Chen2018FEMOSAA,nair2018finding,DBLP:journals/pieee/ChenBY20,DBLP:conf/wosp/0001BWY18,DBLP:journals/csur/ChenBY18}. 
However, 
there are also many other cases, 
such as the above one, 
wherein only the optimization of a single performance attribute is of interest, 
whose minimization (or maximization) serves as the sole performance objective in consideration. 
In another scenario, 
machine learning systems deployed by large organizations 
(e.g., GPT-3~\cite{brown2020language}), 
or those in the health care domain~\cite{DBLP:conf/bcb/AhmadET18}, often concern mainly on the accuracy, 
while caring little about the overhead/resources incurred for training. 
This has been well-echoed from the literature on software configuration tuning, in majority of which only a single performance attribute is considered at a time~\cite{DBLP:journals/jmlr/BergstraB12,DBLP:conf/sigmetrics/YeK03,DBLP:conf/sigsoft/OhBMS17,DBLP:conf/www/XiLRXZ04,DBLP:conf/hpdc/LiZMTZBF14,DBLP:conf/sc/BehzadLHBPAKS13,DBLP:conf/icse/LiX0WT20,DBLP:conf/kbse/LiXCT20}.

\begin{figure}[t!]
	\centering
	\includegraphics[width=0.7\columnwidth]{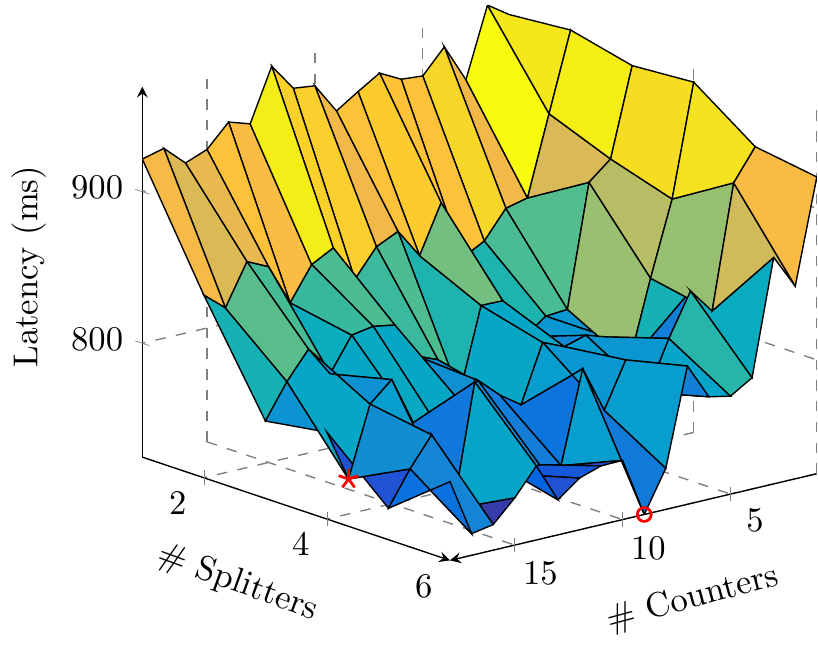}
	\caption{A projected landscape of the performance objective \textit{Latency} 
		with respect to configuration options \texttt{Splitters} and \texttt{Counters} 
		for \textsc{Storm} under the \textsc{WordCount} benchmark.	
		\copy\aMark~is the global optimum and \copy\bMark~is one of the locally optimal latency that an optimizer needs to escape from.	
}
	\label{fig:example}
\end{figure}

Despite only a single performance attribute is of concern,
such an optimization scenario is not easy to deal with. 
This is because (1) the configurable systems involve a daunting number of configuration options with complex interactions, rendering a black-box to the software engineers~\cite{DBLP:conf/sigsoft/XuJFZPT15,DBLP:conf/icse/Chen19b,DBLP:journals/tse/ChenB17};
(2) the number of possible configurations to examine can be high~\cite{DBLP:journals/tsc/ChenB17} and the measurement of each configuration through running the software system is often expensive~\cite{DBLP:conf/mascots/JamshidiC16};
and (3) there is generally a high degree of sparsity in the configurable software systems~\cite{nair2018finding}, 
i.e., the close configurations can also have radically different performance.
The last characteristic poses a particular challenge to any automatic tuning process in finding the optimal configuration (performance), because 
firstly different configurations  
may achieve locally good, but globally undesired performance (e.g., local optima); 
and secondly, the landscape of a (local) optimum's neighborhood can be steep and rugged --- 
if the tuning is trapped in a local optimum, 
it may be hard to escape from it as their neighboring configurations often perform worse than it.
As an example, 
Figure~\ref{fig:example} shows the projected configuration landscape for \textsc{Apache Storm} (2 out of 6 configuration options), 
where it can be clearly seen that even with this simplified version, 
the landscape is rather rugged and contains steep ``local optimum traps'',
resulting in significant difficulty in the tuning.

To address the above challenges,
a number of optimizers from the Search-Based Software Engineering (SBSE) paradigm have been presented, 
such as random search~\cite{DBLP:journals/jmlr/BergstraB12,DBLP:conf/sigmetrics/YeK03,DBLP:conf/sigsoft/OhBMS17}, hill climbing~\cite{DBLP:conf/www/XiLRXZ04,DBLP:conf/hpdc/LiZMTZBF14}, genetic algorithm~\cite{DBLP:conf/sc/BehzadLHBPAKS13,DBLP:conf/sigsoft/ShahbazianKBM20}, and simulated annealing~\cite{DBLP:conf/icpads/DingLQ15,guo2010evaluating}. 
To seek the global optimum (best performance of the concerned performance attribute) while avoiding being trapped in local optima, 
these methods focus on the ``internal'' components of the optimizer.
They work on designing novel search operators (i.e., the way to change the configuration structure, for example, increasing the
neighbourhood size of randomly mutated configurations~\cite{DBLP:conf/sigsoft/OhBMS17}),
or developing various search strategies 
(i.e., the way to balance exploration and exploitation, for example, 
restarting the search in hill climbing~\cite{DBLP:conf/www/XiLRXZ04}). 
However, 
a major limitation of such single-objective optimizers is 
that the goal to find the global optimum is ``less oriented'' 
as there is no clear ``incentive'' to encourage them to traverse the wide search space 
and locating many local optima as possible, 
thus finding the best one in a resource-efficient manner.


In this paper, 
we look to tackle this software configuration tuning problem (with a single performance concern) from a different perspective. 
In contrast to the effort made by the existing works on the development of the optimizer,
we work on the optimization model.
We present a multi-objective optimization model for this single-objective problem, 
to help the search avoid being trapped in local optima and progressively explore the entire objective space --- 
an approach that belongs to the concept called \textbf{multi-objectivization}~\cite{Knowles2001}.

Multi-objectivization, 
which transforms a single-objective optimization problem into a multi-objective one, 
is not particularly unusual in SBSE. 
In several SE scenarios, 
researchers carefully design an auxiliary objective as a helper, 
along with the target objective (i.e., the original objective), 
for a multi-objective optimizer to deal with~\cite{DBLP:journals/tse/YuanB20,derakhshanfar2020good,DBLP:conf/ssbse/MkaouerKBC14,DBLP:conf/ssbse/SoltaniDPDZD18}. 
For example, 
in the crash reproduction problem~\cite{derakhshanfar2020good}, 
a new auxiliary objective was created to check how widely a test case covers the code, 
which is in strong conflict with the target objective that measures how far a test is from the particular line(s)-of-code 
that reproduces the crash.


However, 
a pitfall of this approach is that the auxiliary objective needs a delicate design 
(e.g., to make it rather conflicting with the target objective~\cite{DBLP:conf/ssbse/MkaouerKBC14,derakhshanfar2020good}) 
in order to help the search on the target objective jumps out of local optima. 
The design often requires some similar domain properties between scenarios, 
such as the test cases in the example above, 
which could share some common structures for different software systems at the code level. 
Yet, this assumption does not hold in software configuration tuning, which lies in the configuration level, 
as their configuration options and characteristics can be intrinsically different~\cite{DBLP:conf/sigsoft/XuJFZPT15}, 
while it is difficult to identify the commonality (if any) due to the black-box nature.  

Another drawback of this approach is concerned with its optimization model.
Since the approach treats the two objectives equally during the search, 
solutions that perform well on the auxiliary objective but poorly on the target objective 
will still be regarded as ``optimal'' 
(in the sense of Pareto optimality; see Section~\ref{sec:method}), 
thus being preserved, exploited, and explored repetitively during the search process. 
However, 
such solutions are meaningless to the considered optimization problem;
keeping exploiting them can 
cause  waste of resources (search budget), 
which eventually lowers the chances of finding a better target objective.

In this work, 
we propose a different multi-objectivization model, which 
contains two \textbf{meta-objectives} to optimize 
(hence called meta multi-objectivization model, or MMO;
in contrast, the preceding model which directly optimizes the target and auxiliary objectives is called plain multi-objectivization, or PMO). 
Each of the two meta-objectives has two components.
The first component of both meta-objectives is the target performance objective (e.g., latency), 
thereby only those configurations that perform well on the target being in favor. 
The second component, 
which is related to the other given auxiliary performance objective (e.g., throughput, based on whatever that is available), 
is a completely conflicting term for the two meta-objectives. 
The reason for this design is that 
we hope to keep the target performance objective as a primary term in the model to preserve the tendency towards its optimality, 
and at the same time, 
we want the configurations with different values on the auxiliary performance objective to be incomparable.
We are not interested in minimizing/maximizing the auxiliary performance objective 
since we do not know which value of it can lead to the best result on the target performance objective,
but we wish to keep a good amount of configurations with various values of the auxiliary performance objective in the search, 
thus not being trapped in local optima (we will elaborate this in Section~\ref{sec:method}).

It is worth mentioning that software configuration tuning provides a well-fitting avenue for multi-objectivization:
similar kind of configurable software systems would inherently come with at least two prevalent performance attributes, 
e.g., the latency and throughput for stream processing systems~\cite{nair2018finding}; 
the accuracy and training/inference time for machine learning systems~\cite{DBLP:conf/nips/SculleyHGDPECYC15}.
Since we run the software system in the tuning anyway, 
one would merely need to measure how the configurations affect at least one other performance attribute, 
using penalty of readily available tools/API~\cite{bogetoft2013performance}. 
Such an attribute can then contribute to the auxiliary objective in multi-objectivization without the need for a specific design.

\vspace{3pt}
Overall, 
the contributions of this work are:

\begin{itemize}
    \item Unlike existing work for the software configuration tuning which puts efforts on 
    the ``internal part'' of the optimization (i.e., improving the search operators of various optimizers), 
    we work on the ``external part'' --- 
    multi-objectivizing this single-objective optimization scenario.
    
    \item We present a meta multi-objectivization model, MMO, as opposed to the existing multi-objectivization model 
    considered in other SBSE scenarios which directly optimizes the target and auxiliary objectives simultaneously (i.e., PMO). 
    We show, analytically and experimentally, 
    why MMO is more suitable than PMO for software configuration tuning.
   
    \item We conduct extensive experiments on eight commonly used real-world software systems/environments 
    that are of diverse domains, scales, settings, search space, and performance attributes.
    Equipped with a classic multi-objective optimizer, NSGA-II~\cite{Deb2002}, 
    we compare our model with four state-of-the-art single-objective optimizers that underpin many prior works on software configuration tuning, 
    i.e., random search with high neighbourhood radius~\cite{DBLP:journals/jmlr/BergstraB12,DBLP:conf/sigmetrics/YeK03,DBLP:conf/sigsoft/OhBMS17}, stochastic hill climbing with restart~\cite{DBLP:conf/www/XiLRXZ04,DBLP:conf/hpdc/LiZMTZBF14}, single-objective genetic algorithm~\cite{DBLP:conf/sc/BehzadLHBPAKS13,DBLP:conf/sigsoft/ShahbazianKBM20}, and simulated annealing~\cite{DBLP:conf/icpads/DingLQ15,guo2010evaluating}. 
    
    \item We investigate three different instances of MMO 
    and their sensitivity to a critical internal parameter in the model.   

\end{itemize}

The experiment results are encouraging. 
We show that the proposed MMO model,
compared with the best state-of-the-art single-objective optimizer, 
achieves better result (up to 42\% gain, with statistical significance and non-trivial effect sizes) on the target performance objective for the majority of the cases,
while generally consuming less resources (number of measurements that reflects the time and computation needed) as low as 24\%.
This contrasts with the PMO model which in general performs worse than the best single-objective optimizer.
We can conclude that our model: 
\begin{itemize}
\item is generally safe and effective to use, 
while exhibiting marginal differences between different model instances;
\item is overall resource-efficient, 
meaning that it is suitable for expensive problems like software configuration tuning; 
\item may be sensitive to its parameter setting, 
however, 
there exist some good ``rule-of-the-thumb'' values across the cases.

\end{itemize}

All source code and data can be accessed at our GitHub repository: \href{https://github.com/taochen/mmo-fse-2021}{\texttt{\textcolor{blue}{https://github.com/taochen/mmo-fse-2021}}}.

The rest of this paper is organized as follows. 
Section~\ref{sec:prob} introduces some background information. 
Section~\ref{sec:method} elaborates the design of our meta multi-objectivization model. 
Section~\ref{sec:exp} presents our experiment methodology, 
followed by a detailed discussion of the results in Section~\ref{sec:result}. 
The usefulness of the proposed model and threats to validity are discussed in Section~\ref{sec:discussion}. 
Sections~\ref{sec:related} and~\ref{sec:con} analyze the related work and conclude the paper, respectively.


%% file: problem.tex
\section{Preliminaries}
\label{sec:prob}

In this section, we describe the necessary background.

\subsection{Software Configuration Tuning Problem}


A configurable software system often comes with a set of critical configuration options such that the $i$th option is denoted as $x_i$, 
which can be either a binary or integer variable,
where $n$ is the total number of options.
The search space, 
$\mathbfcal{X}$, 
is the Cartesian product of the possible values for all the $x_i$. 
Formally, 
when only a single performance concern is of interest (such as latency, throughput, or accuracy), 
the goal of software configuration tuning is to achieve\footnote{Without loss of generality, we assume minimizing the performance objective.}:
\begin{align}
	\argmin~f(\vect{x}),~~\vect{x} \in \mathbfcal{X}
	\label{Eq:SOP}
\end{align}
where $\vect{x} = (x_1, x_2, ..., x_n)$.
This is a classic \textit{single-objective optimization model} and the measurement of $f$ is entirely case-dependent according to the target software and the corresponding performance attribute; 
thus we make no assumption about its characteristics.

\subsection{Multi-Objectivization}

Multi-objectivization is the process of transforming a single-objective optimization problem 
into a multi-objective one, 
in order to make the search easier to find the global optimum.
It can be realized by adding a new objective (or several objectives) to the original objective 
or replacing the original objective with a set of objectives. 
The motivation is that 
since in complex problem landscape, 
the search may get trapped in local optima when considering the original objective 
(due to the total order relation between solutions on the objective), 
considering multiple objectives may make similarly-performed solutions incomparable 
(i.e., Pareto nondominated to each other),
thus helping the search jump out of local optima~\cite{Knowles2001}.

Two solutions being Pareto nondominated means 
that one is better than the other on some objective and worse on some other objective. 
Formally,
for two solutions $\vect{x}$ and $\vect{y}$, 
we call $\vect{x}$ and $\vect{y}$ nondominated to each other 
if $\vect{x} \nprec \mathbf{y} \wedge \vect{y} \nprec \vect{x}$,
where $\nprec$ is the negation of ``to Pareto dominate'' ($\prec$), 
the superiority relation between solutions for multi-objective optimization.
That is, 
considering a minimization problem with $m$ objectives,
$\vect{x}$ is said to \textit{(Pareto) dominate} $\vect{y}$ 
(denoted as $\vect{x}\prec \vect{y}$) 
if $f_i(\vect{x}) \leq f_i(\vect{y})$ for $1 \leq i \leq m$ and 
there exists at least one objective $j$ on which $f_j(\vect{x}) < f_j(\vect{y})$.
Pareto dominance is a partial order relation, 
and thus there typically exist multiple optimal solutions in multi-objective optimization.
For a solution set $\vect{X}$,
a solution $\vect{x} \in \vect{X}$ is called \textit{Pareto optimal} to $\vect{X}$
if there is no solution $\in \vect{X}$ that dominates $\vect{x}$. 
When $\vect{X}$ is the collection of all feasible solutions for a multi-objective problem,
$\vect{x}$ becomes an optimal solution to the problem, 
and the set of all Pareto optimal solutions of the problem is called its \textit{Pareto optimal set}. 

Multi-objectivization is not uncommon in the modern optimization realm, 
particularly to the evolutionary computation community~\cite{Knowles2001,Cai2006,Ishibuchi2007,Song2014,Steinhoff2020}.
To tackle various challenging single-objective optimization problems, 
researchers put much effort in introducing/designing additional objectives,
e.g., creating sub-problems (sub-objectives) of the original objective~\cite{Knowles2001}, 
converting the constraints into an additional objective~\cite{Cai2006},
constructing similar adjustable objectives~\cite{Ishibuchi2007},
considering one of the decision variables~\cite{Song2014},
or even adding a man-made less relevant objective function~\cite{Steinhoff2020}. 





%% file: method.tex
\section{Multi-Objectivization in Software Configuration Tuning}
\label{sec:method}

Here we present the designs of our MOO model and how they are derived from the key properties in software configuration tuning.

\subsection{Key Properties in Configuration Tuning}

We observed that, in general, software configuration tuning bears the following properties.

\underline{\textbf{Property 1:}} As shown in Figure~\ref{fig:example} and what has already been reported~\cite{nair2018finding,DBLP:conf/mascots/JamshidiC16}, the configuration landscape for most configurable software systems are rather rugged with numerous local optima at varying slopes. 
Therefore the tuning, once the search is trapped at a local optimum, would be difficult to progress. 
This is because if only the concerned performance attribute is used to guide the search, 
and all the surrounding configurations on a local optimum are significantly inferior to it, 
then the search focus would have no much drive to move away from that local optimum. 
As a result, 
a good optimization model has additional ``tricks'' to avoid comparing configurations solely based on the single performance attribute. 

\underline{\textbf{Property 2:}} A single measurement of configuration is often expensive. For example, Valov et al.~\cite{DBLP:conf/wosp/ValovPGFC17} reported that sampling all values of 11 configuration options for \textsc{x264} needs 1,536 hours. This means that the resource (search budget) in software configuration tuning is highly valuable, hence utilizing them efficiently is critical.

\begin{figure}[t!]
	\centering
	\includegraphics[width=\columnwidth]{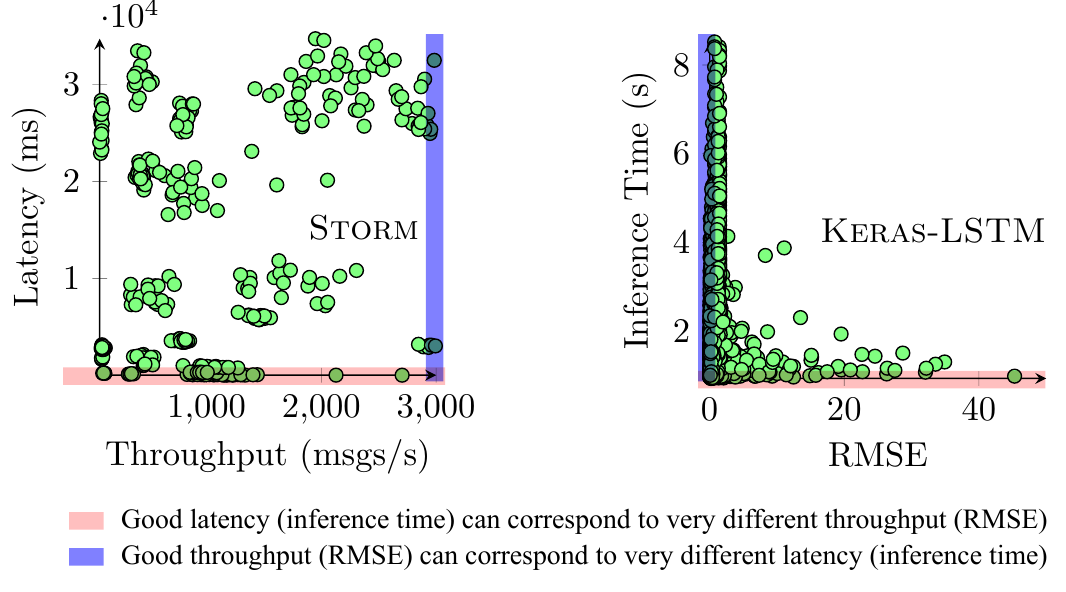}
	\caption{Measured configurations for \textsc{Storm} and \textsc{Keras-LSTM}. The points that Property 3 refers to are highlighted: very good or bad results on one performance objective can both correspond to similarly good value on the other.}
	\label{fig:prop3}
\end{figure}

\underline{\textbf{Property 3:}} The correlation between different performance attributes is often uncertain, 
as different configurations may have different effects on distinct attributes.
As such, we observed that the configurations may achieve extremely good or bad performance on one while having similarly good results on the other, as illustrated in Figure~\ref{fig:prop3}. The reasons for this can vary. Taking the \textsc{Storm} from Figure~\ref{fig:prop3} (left) as an example; suppose that in a multi-threaded and multi-core environment with 100 successful messages, if a configuration $\vect{A}$ enables each of these messages to be processed at 30ms, then the latency and throughput are ${{100 \times 30} \over {100}} = 30$ms and ${100 \over 30} = 3.33$ msgs/ms, respectively. 
In contrast, another configuration $\vect{B}$ may restrict the parallelism (e.g., lower \texttt{spout\_num}), hence there could be 50 messages processed at 20ms each\footnote{The relief of peak CPU load could allow the process of each message faster.} while the other 50 are handled at 40ms each (including 20ms queuing time due to reduced parallelism). Here, the latency remains at ${{50 \times 20 + 50 \times 40} \over {100}} = 30$ms but the throughput is changed to ${100 \over 40} = 2.5$ msgs/ms, which is a 25\% drop. Therefore, we should not presume either a strict conflicting or harmonic correlation between the performance attributes.


As such, 
a good optimization model for software configuration tuning should take the above properties into account.

\subsection{Plain Multi-Objectivization Model (PMO)}

A straightforward idea to perform multi-objectivization is to add an auxiliary objective to optimize, 
along with the target objective.
This is what has been commonly used in other SBSE scenarios (e.g.,~\cite{derakhshanfar2020good}).
That PMO model can be formulated as:
\begin{align}
\begin{split}
\text{minimize}
\begin{cases}
f_t(\vect{x})\\
f_a(\vect{x})\\
\end{cases}
\end{split}
\label{Eq:pmo_model}
\end{align}
where $f_t(\vect{x})$ denotes the target performance objective (i.e., the concerned one) 
and $f_a(\vect{x})$ denotes the auxiliary performance objective\footnote{Without loss of generality, 
	 we use the minimization form of the auxiliary performance objective; 
	the maximization ones can be trivially converted, e.g., by multiplying $-1$.}.

Putting in the context of software configuration tuning, 
the PMO model may cover \textbf{Property~1}, 
because the natural Pareto relation ensures that the target performance objective is no longer a sole indicator to guide the search. However, 
it does not fit \textbf{Property~2} as PMO additionally optimizes the auxiliary performance objective. 
As such, 
configurations that perform well on the auxiliary performance objective but poorly on the target performance objective
are still regarded as optimal in PMO, 
despite being meaningless to the original problem. 
This can result in a significant waste of resources. 
In addition,
PMO does not consider \textbf{Property~3} as it often assumes conflicting correlation between the two objectives~\cite{DBLP:conf/ssbse/MkaouerKBC14,derakhshanfar2020good}, 
which is hard to assure in software configuration tuning.


\subsection{Our Meta Multi-Objectivization Model}

Unlike PMO, 
our meta multi-objectivization (MMO) model creates two meta-objectives based on the performance attributes. 
The aim is to drive the search towards the optimum of the target performance objective, 
and at the same time, not being trapped in local optima. 
Specifically,
we want to achieve two goals:

\begin{itemize}
    \item[---] \textbf{Goal 1:} optimizing the target performance objective still plays a primary role, 
    thus no resource waste on, for example, optimizing the auxiliary one (this fits in \textbf{Property 2});

    \item[---] \textbf{Goal 2:} but those with different values of the auxiliary performance objective are more likely to be incomparable (i.e., Pareto nondominated), 
    thus the search would not be trapped in local optima (this relates to \textbf{Properties 1} and \textbf{3}).
\end{itemize}



Formally, 
the proposed model with two meta-objectives $g_1(\vect{x})$ and $g_2(\vect{x})$ is constructed as:
\begin{align}
\begin{split}
\text{minimize}
\begin{cases}
g_1(\vect{x}) = f_t(\vect{x}) + \varphi(f_a(\vect{x}))\\
g_2(\vect{x}) = f_t(\vect{x}) - \varphi(f_a(\vect{x}))\\
\end{cases}
\end{split}
\label{Eq:model}
\end{align}
whereby
each of the two meta-objectives shares the same target performance objective $f_t(\vect{x})$, 
but differs (effectively being opposite) 
regarding the auxiliary performance objective $f_a(\vect{x})$. 
$\varphi()$ is a composite function that balances the $f_t(\vect{x})$ and $f_a(\vect{x})$. 
In theory, 
the MMO model is generic and hence $\varphi()$ can take different forms to implement specific instances of the model. 
Here we consider its simplest instance $\varphi(f_a(\vect{x})) = wf_a(\vect{x})$
(we will investigate other instances in Section~\ref{sec:exp}). 
That is,
\begin{align}
\begin{split}
\text{minimize}
\begin{cases}
g_1(\vect{x}) = f_t(\vect{x}) + wf_a(\vect{x})\\
g_2(\vect{x}) = f_t(\vect{x}) - wf_a(\vect{x})\\
\end{cases}
\end{split}
\label{Eq:model2}
\end{align}
where $w$ is a weight parameter that allows fine-tuning of the balance; 
different systems may need different settings. 
Note that in the MMO model,
both the target and the auxiliary performance objectives need to be normalized for commensurability. 



\begin{figure}[tbp]
	\begin{center}
 	\setlength{\tabcolsep}{0.2mm}
		\footnotesize
		\begin{tabular}{@{}cc@{}}
			\vspace{-1pt}
			\includegraphics[scale=0.25]{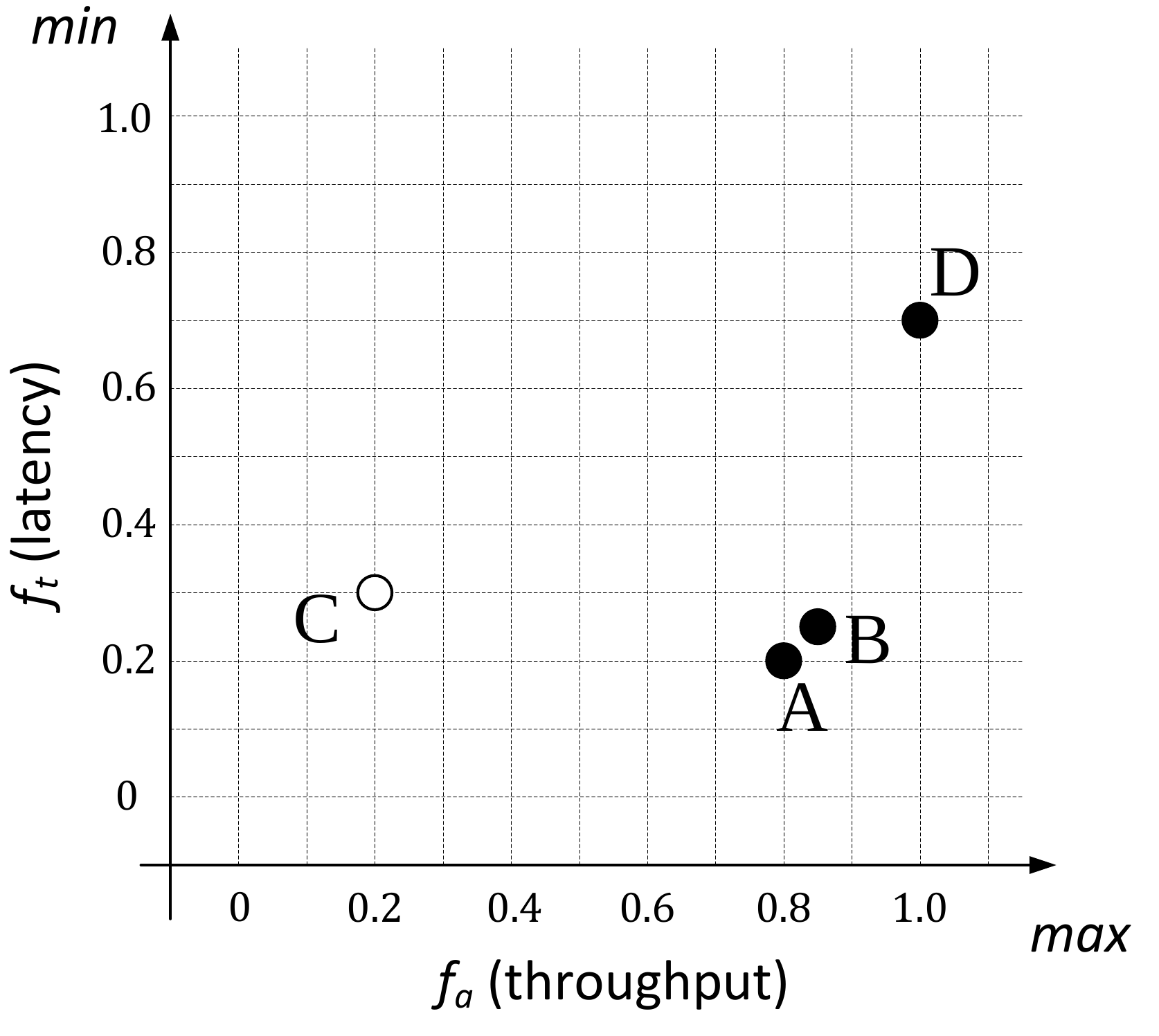}&
			\includegraphics[scale=0.25]{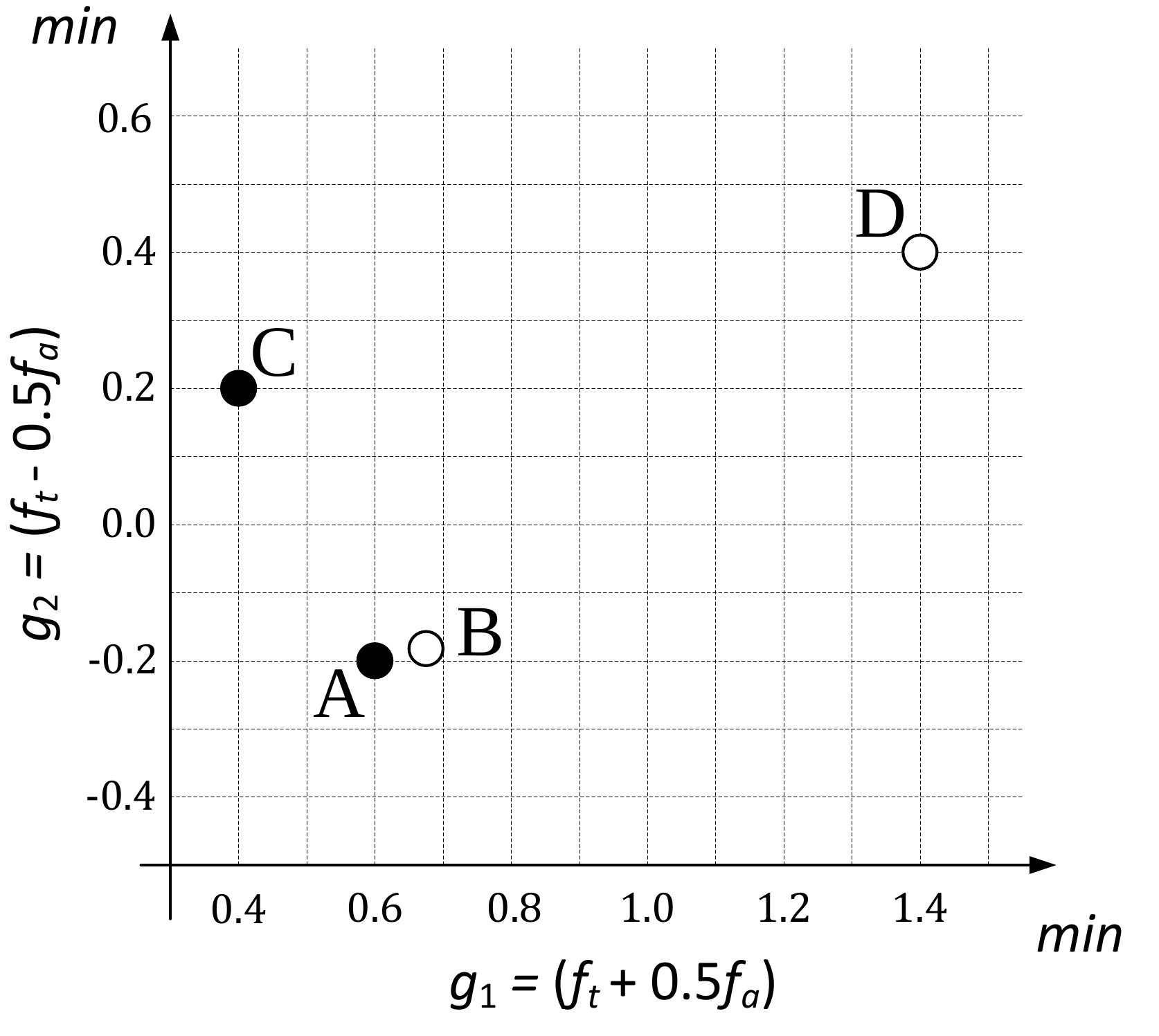}\\
			(a) The original target-auxiliary space & (b) Our meta bi-objective space \\
		\end{tabular}
	\end{center}
	\vspace{0pt}
	\caption{\small{An illustration of comparison between (a) the PMO model and (b) our MMO model (with the instance of Equation~\ref{Eq:model2}) on \textsc{Storm},
			where the target performance objective is latency (to minimize) and 
			the auxiliary performance objective is throughput (to maximize). 
			Both of them are normalized and the weight is $0.5$ in MMO.
			Let us say $\vect{A}$, $\vect{B}$, $\vect{C}$ and $\vect{D}$ be a set of four configurations to be selected by the two models. 
			The solid circle means the configuration being Pareto optimal to the set. 
			Since the PMO model directly minimizes latency and maximizes throughput (Figure~\ref{Fig:model}a), 
			configurations $\vect{A}$, $\vect{B}$, and $\vect{D}$ are Pareto optimal. 
			However, $\vect{D}$ performs very poorly on latency; 
			preserving it during the search is a waste of resources. 
			In contrast, 
			in our MMO model (Figure~\ref{Fig:model}b), 
			configurations $\vect{A}$ and $\vect{C}$ are Pareto optimal.
			Now comparing configurations $\vect{C}$ with $\vect{B}$,
			since $\vect{C}$ and $\vect{A}$ behave much more differently than $\vect{B}$ and $\vect{A}$ on the throughput,
			it is often more likely that they are located on distant regions in the configuration landscape; 
			thus preserving $\vect{C}$ rather than $\vect{B}$ (in case $\vect{A}$ is preserved) 
			is generally more likely to avoid the search being trapped in the local optimum.}  
	}
	\label{Fig:model}
\end{figure}

To understand the proposed MMO model, 
Figure~\ref{Fig:model} gives an example of \textsc{Storm} on how it distinguishes between different configurations, 
in comparison with the PMO model,
when using latency as the target performance objective $f_t$ and throughput as the auxiliary performance objective $f_a$. 
Suppose that there is a set of four configurations $\vect{A}$, $\vect{B}$, $\vect{C}$ and $\vect{D}$.
Let us say if we want to choose two of them based on their fitness 
(e.g., in order to put some promising configurations into the next-generation population).
For the PMO model (Figure~\ref{Fig:model}a) that minimizes latency and maximizes throughput,
the configuration $\vect{D}$, which performs extremely poor on latency, will certainly be chosen by any multi-objective optimizer,
since it is Pareto optimal and also less crowded than the other Pareto optimal configuration $\vect{A}$ and $\vect{B}$.
In contrast,
for our MMO model (Figure~\ref{Fig:model}b) which minimizes the two meta objectives,
the two configurations that will be chosen are $\vect{A}$ and $\vect{C}$
since they are the only two Pareto optimal ones. 

It is worth noting that for the single-objective optimization model (which only considers latency), 
the two chosen configurations will be $\vect{A}$ and $\vect{B}$.
However,
since $\vect{C}$ and $\vect{A}$ behave much more differently than $\vect{B}$ and $\vect{A}$ on the throughput, 
it is often more likely that they are located in distant regions in the configuration landscape; 
thus preserving $\vect{C}$ rather than $\vect{B}$ (when $\vect{A}$ is preserved) 
is generally more likely to help the search to escape from the local optimum.

By further help to grasp the characteristics of the MMO model, 
we provide five remarks below. 
Remarks 1--3 are related to 
why the target performance objective remains primary in the model (\textbf{Goal~1}).
Remarks 4 and 5 show how it helps to escape local optima via creating ``incomparability'' 
between the configurations with dissimilar values on the auxiliary performance objective (\textbf{Goal 2}).


	\underline{\textbf{Remark 1:}} The global optimum of the original single-objective problem 
	(i.e., the configuration with the best target performance objective)
	is Pareto optimal in MMO 
	(e.g., the configuration $\vect{A}$ in the example of Figure~\ref{Fig:model}).
	This can be immediately obtained by contradiction from Equation~(\ref{Eq:model}).
	
	\underline{\textbf{Remark 2:}} A similar but more general observation is 
	that a configuration will never be dominated by another that has a worse target performance objective.
	This can also be derived from Equation~(\ref{Eq:model}) --- 
	If configuration $\vect{x_1}$ has a better target performance objective than $\vect{x_2}$ 
	(i.e., $f_t(\vect{x_1}) < f_t(\vect{x_2})$), 
	then whatever their auxiliary performance objective values are,
	$\vect{x_2}$ will not be better than $\vect{x_1}$ on both $g_1$ and $g_2$;
	in the best case for $\vect{x_2}$, 
	they are nondominated to each other 
	(e.g., the configuration $\vect{B}$ versus $\vect{C}$ in Figure~\ref{Fig:model}).
	
	\underline{\textbf{Remark 3:}} The above two remarks apply to the target performance objective, 
	but not to the auxiliary performance objective. 
	This is a key difference from the PMO model, 
	where both objectives are subject to these remarks;
	thus the configuration $\vect{D}$ in the example of Figure~\ref{Fig:model}, 
	which is meaningless to the original problem, 
	is treated as being optimal in PMO but not in MMO.

	\underline{\textbf{Remark 4:}} MMO does not bias to a higher or lower value on the auxiliary performance objective, 
	in contrast to PMO.
	This makes sense since, 
	as explained in \textbf{Property 3}, 
	we do not know for certain that what value of the auxiliary performance objective corresponds to the
	best target performance objective.

	\underline{\textbf{Remark 5:}} 
	Configurations with dissimilar auxiliary performance objective values tend to be incomparable
	(i.e., nondominated to each other) 
	even if one is
	fairly inferior to the other on the target performance objective.
	For example,
	the configuration $\vect{C}$ in Figure~\ref{Fig:model},
	which has worse latency than $\vect{A}$,
	is not dominated by $\vect{A}$ as their throughput are rather different.
	In contrast, 
	the configuration $\vect{B}$,
	which even has better latency than $\vect{C}$,
	is dominated by $\vect{A}$, 
	as they are similar on  throughput. 
	This enables the model to keep exploring diverse promising configurations during the search, 
	thereby a higher chance to find the global optimum.

%% file: study.tex
\section{Experimental Setup}
\label{sec:exp}

In this section, we articulate the experimental methodology for evaluating our MMO model and its instances.

\subsection{Research Questions}
\label{sec:rq}


Our experiment investigates the following research questions (RQs):

\begin{itemize}

    \item[---] \textbf{RQ1:} How effective is the MMO model?
    \item[---] \textbf{RQ2:} How resource-efficient is the MMO model?
    \item[---] \textbf{RQ3:} What is the sensitivity of the MMO model to its weight?
\end{itemize}

We ask \textbf{RQ1} to verify whether our MMO model can better help to overcome the issue of local optima, i.e., by providing better results than the single-objective counterpart and PMO under the same search budget. We investigate \textbf{RQ2} to examine whether the resources (the number of measurements) are consumed to reach a certain level of performance in a reasonably efficient manner. We use \textbf{RQ3} to study whether the weight in MMO is critical. 


\input{tables/sys}



\subsection{Subject Software Systems}

As shown in Table~\ref{tb:sys}, we experiment on a set of commonly used real-world software systems and environments~\cite{nair2018finding,DBLP:conf/mascots/JamshidiC16,DBLP:conf/mascots/MendesCRG20,DBLP:conf/sigsoft/JamshidiVKS18}, whose single measurement is expensive\footnote{To ensure robustness, each measurement consists of 5 repeated samples and the median value is used.}. They come from diverse domains, e.g., stream processing (SP) and deep learning (DL), while having different performance attributes, scale, and search space. 


Each software system comes with two performance objectives, which are chosen arbitrarily from prior work~\cite{nair2018finding,DBLP:conf/mascots/JamshidiC16,DBLP:conf/mascots/MendesCRG20,DBLP:conf/sigsoft/JamshidiVKS18}. In all experiments, we use each of their two performance attributes as the target performance objective in turn while the other serves as the auxiliary performance objective. 




We use the same configuration options and their ranges as studied in the prior work~\cite{nair2018finding,DBLP:conf/mascots/JamshidiC16,DBLP:conf/mascots/MendesCRG20,DBLP:conf/sigsoft/JamshidiVKS18}, since those have been shown to be the key ones for the software systems under the related environment. As a result, although some subject software appears to be the same, their actual search spaces are different, such as \textsc{Storm/WC} and \textsc{Storm/RS}.




\subsection{Tuning Settings}
\label{sec:settings}

\subsubsection{Models, MMO Instances and Optimizers}

For the single-objec-tive optimization model, we examine four state-of-the-art optimizers that are widely used in software configuration tuning, all of which deal with local optima in different ways:

\begin{itemize}
    \item Random Search (RS) with a high neighbourhood radius to escape from the local optima~\cite{DBLP:journals/jmlr/BergstraB12,DBLP:conf/sigmetrics/YeK03,DBLP:conf/sigsoft/OhBMS17}.
    \item Stochastic Hill Climbing with restart (SHC-r)~\cite{DBLP:conf/www/XiLRXZ04,DBLP:conf/hpdc/LiZMTZBF14}, aiming to avoid local optima by using different starting points.
    \item Single-Objective Genetic Algorithm (SOGA)~\cite{DBLP:conf/sc/BehzadLHBPAKS13,DBLP:conf/sigsoft/ShahbazianKBM20} that seeks to escape local optima by using variation operators.
    \item Simulated Annealing (SA)~\cite{DBLP:conf/icpads/DingLQ15,guo2010evaluating} that tackles local optima by stochastically accepting inferior configurations. 
\end{itemize}




Recall from Equation (2), our MMO model can be instantiated in different forms.
In the experiments, 
we consider three alternatives:

\begin{itemize}
	\item[---] \textbf{{MMO-Linear}:} $\varphi(f_a(\vect{x})) = wf_a(\vect{x})$. 
	\item[---] \textbf{{MMO-Sqrt}:} $\varphi(f_a(\vect{x})) = w\sqrt{f_a(\vect{x})}$.
	\item[---] \textbf{{MMO-Square}:} $\varphi(f_a(\vect{x})) = wf_a^2(\vect{x})$.
\end{itemize}


We examine all the above instances of the MMO model, together with the PMO. 
While our model does not tie to any specific multi-objective optimizer, 
we use NSGA-II for both MMO and PMO in this work, 
because (1) it has been predominately used for software configuration tuning in prior work when multiple performance attributes are of interest~\cite{Chen2018FEMOSAA,DBLP:conf/wosp/SinghBSH16,DBLP:journals/infsof/ChenLY19,DBLP:conf/icpads/KumarBCLB18,DBLP:journals/jss/SobhyMBCK20}; (2) it shares many similarities with the SOGA that we compare in this work.
Note that MMO may not be able to work with some multi-objective optimizers specifically designed for SBSE problems 
where the objectives are not treated equally, 
such as \cite{Panichella2015,Hierons2016,Hierons2020}.



All those optimizers can, but do not have to, rely on a surrogate. Since we focus on the optimization model, the ability to omit the surrogate model is desirable, as it has been shown that such a surrogate can be highly inaccurate~\cite{DBLP:conf/cloud/ZhuLGBMLSY17} and hence creates noises in our experiments. 
In this work, all optimization models and optimizers are implemented in Java, using jMetal~\cite{DBLP:journals/aes/DurilloN11} and Opt4J~\cite{DBLP:conf/gecco/LukasiewyczGRT11}.

\subsubsection{Weight Values}

In our experiments, we evaluate a set of weight values, i.e., $w \in \{0.01,0.1,0.3,0.5,0.7,0.9,10\}$, for all MMO instances. Those are merely pragmatic settings without any sophisticated reasoning. In this way, we aim to examine whether the MMO model can be effective by choosing from some randomly given weight values. To make the performance objectives commensurable in MMO, we use \textit{max-min} scaling~\cite{dodge2006oxford}. However, since the bounds are often unknown, we update them dynamically as the tuning proceeds; this is a widely used approach in SBSE~\cite{DBLP:conf/sigsoft/ShahbazianKBM20}. 



\subsubsection{Search Budget}

Since the measurement is expensive, we repeat all experiments 30 runs with a search budget of 2 hours each, as suggested in prior work~\cite{DBLP:conf/mascots/JamshidiC16}. However, directly using the time as a termination criterion would cause the search to suffer non-trivial interference given the number of experiments we need to run in parallel. To avoid this, for each software system, we incrementally (100 each step) measured distinct configurations on a dedicated machine using random sampling until the time budget is exhausted. In this way, we collect the number of measurements (the median of 5 repeats), as shown in Table~\ref{tb:settings}, that serve as the termination criterion for the configuration tuning thereafter.

Since the search budget reflects the number of measurements permitted in 2 hours, in each run, we cached the measurement of every distinct configuration, which can be reused directly when the same configuration appears again during the search. In other words, only the distinct configurations would consume the budget. 





\subsubsection{Other Parameters}

For SOGA and NSGA-II, we apply the binary tournament for mating selection, together with the boundary mutation and uniformed crossover, as used in prior work~\cite{Chen2018FEMOSAA,DBLP:conf/sigsoft/ShahbazianKBM20,DBLP:journals/infsof/ChenLY19}. The mutation and crossover rates are set to 0.1 and 0.9, respectively, as commonly set in software configuration tuning~\cite{Chen2018FEMOSAA}.

However, what we could not decide easily is the population size for SOGA and NSGA-II. Therefore, for each software system, we examine different population sizes, i.e., $\{10,20,...,100\}$ in preliminary runs. We used the largest size that enables the population change to be less than 10\% in the last 10\% of the generations over both optimizers, performance objectives, and weights. The results are shown in Table~\ref{tb:settings}. In this way, we seek to reach a good balance between convergence (smaller population change) and diversity (larger population size) under a search budget.

    
    



\input{tables/alg}


\subsection{Comparison and Statistical Test}

\subsubsection{Metric}

Since only the target objective is of interest, 
we do not need to consider the quality of the auxiliary objective~\cite{Li2020}.
We use the average normalized percentage gain~\cite{hake1998interactive} of the target objective on the MMO (or PMO) model against on the single-objective counterpart\footnote{We convert all maximizing objectives by multiplying $-1$.}, 
which is defined as:
\begin{equation}
\text{Normalized \% Gain} =
{{1\over n} \times \sum^n_{i=1} {{y_i - x_i} \over {y_i - y_o}}} \times 100
\end{equation}
whereby $x_i$ and $y_i$ are the objective value of the single performance concern at the $i$th run for a multi-objectivization model and the best (average) single-objective counterpart, respectively. $y_o$ is an utopian performance that none of the optimizers can achieve. In this work, we set $y_o=v_o-q$ wherein $v_o$ is the optimal performance value found from all optimizers; and $q$ is the distance of the closest sample $s$ to $v_o$ over all cases, such that $s \neq v_o$\footnote{We found that for all software systems studied in this work, there exist $q < v_o$.}. Clearly, when the normalized \% gain is zero or negative, it implies that the multi-objectivization model is similar or even worse off, respectively. Note that the objective values are sorted for a total of $n$ runs where $n=30$. According to Hake~\cite{hake1998interactive}, the normalized \% gain is a more suitable metric than its non-normalized version (without $y_o$) because:


\begin{itemize}
    \item It has been used as a standard metric in many domains~\cite{hake1998interactive}.
    \item It can more accurately capture the spread~\cite{marx2007normalized}.
    \item More importantly, it rewards (or penalizes) improvement (or degradation) more when the $y_i$ is closer to the (approximately) optimal value. For example, improving the latency from 100s to 50s shares the same non-normalized \% gain as from 50s to 25s (i.e., 50\%). However, given the severe issue of local optima in software configuration tuning, the latter case can be much more difficult to achieve than the former and hence deserves a greater reward. Suppose that the utopian performance is 20s in the above example, the normalized \% gain for the two cases would be 62.5\% and 83.3\%, respectively.
\end{itemize}

\subsubsection{Statistical Methods}

We use the following statistical methods: 


\begin{itemize}

    \item[---]\textbf{Wilcoxon signed-rank test~\cite{Wilcoxon1945IndividualCB}:} We apply this with $a=0.05$ to investigate the statistical significance of the performance objective comparisons over all 30 runs, as it is a non-parametric statistical test and has been recommended in software engineering research for its strong statistical power on pair-wise comparisons~\cite{ArcuriB11}. If the $p<0.05$, we say the magnitude of differences in the comparisons are significant.
    
    
    \item[---]\textbf{$\mathbf{\hat{A}_{12}}$ effect size~\cite{Vargha2000ACA}:} We use $\hat{A}_{12}$ to verify the effect size over 30 runs. When comparing a multi-objectivization model and its single-objective counterpart in this work, $\hat{A}_{12}>0.5$ denotes that the multi-objectivization is better for more than 50\% of the times. In particular, $0.56\leq \hat{A}_{12}<0.64$ indicates a small effect size while $0.64 \leq \hat{A}_{12} < 0.71$ and $\hat{A}_{12} \geq 0.71$ mean a medium and a large effect size, respectively.

\end{itemize}

%% file: tables/sys.tex
\begin{table}[t!]
\caption{Configurable software systems studied.}
\label{tb:sys}
\setlength{\tabcolsep}{0.6mm}
\centering
\footnotesize
\begin{tabular}{lllcl}\toprule

\textbf{Software}&\textbf{Domain}&\textbf{Performance Objective}&\textbf{$\lvert \mathbfcal{O} \rvert$}&\textbf{Search Space}\\

\midrule

\textsc{Trimesh}&Mesh&\textsc{O1:} \# Iteration; \textsc{O2:} Latency&13&239,260\\


\rowcolor{steel!10}\textsc{x264}&Video&\textsc{O1:} PSNR; \textsc{O2:} Energy Usage&17&53,662\\

\textsc{Storm/WC}&SP&\textsc{O1:} Throughput; \textsc{O2:} Latency&6&2,880\\

\rowcolor{steel!10}\textsc{Storm/RS}&SP&\textsc{O1:} Throughput; \textsc{O2:} Latency&6&3,839\\

\textsc{Storm/SOL}&SP&\textsc{O1:} Throughput; \textsc{O2:} Latency&13&2,048\\

\rowcolor{steel!10}\textsc{Keras-DNN/DSR}&DL&\textsc{O1:} AUC; \textsc{O2:} Inference Time&13&3.32$\times 10^{13}$\\

\textsc{Keras-DNN/Coffee}&DL&\textsc{O1:} AUC; \textsc{O2:} Inference Time&13&2.66$\times 10^{13}$\\

\rowcolor{steel!10}\textsc{Keras-LSTM}&DL&\textsc{O1:} RMSE; \textsc{O2:} Inference Time&13&7,040\\


\bottomrule
\end{tabular}
\centering
 \begin{tablenotes}
    \footnotesize
    \item $\lvert \mathbfcal{O} \rvert$ denotes number of options. We run all systems under their standard benchmarks. More details can be found at: \href{https://github.com/taochen/mmo-fse-2021}{\texttt{\textcolor{blue}{https://github.com/taochen/mmo-fse-2021}}}.
    \end{tablenotes}
\end{table}

%% file: tables/alg.tex
\begin{table}[t!]
\caption{Population size and measurement search budget.}
\label{tb:settings}
\setlength{\tabcolsep}{0.9mm}
\footnotesize
\centering
\begin{tabular}{ccc|ccc}\toprule

\textbf{Software}&\textbf{Pop. Size}&\textbf{Budget}&\textbf{Software}&\textbf{Pop. Size}&\textbf{Budget}\\
\midrule

\textsc{Trimesh}&20&1000&\textsc{x264}&50&2,500\\
\rowcolor{steel!10}\textsc{Storm/WC}&50&600&\textsc{Storm/RS}&50&900\\
\textsc{Storm/SOL}&50&700&\textsc{Keras-DNN/DSR}&60&800\\
\rowcolor{steel!10}\textsc{Keras-DNN/Coffee}&50&900&\textsc{Keras-LSTM}&20&400\\


\bottomrule
\end{tabular}
\end{table}

%% file: results.tex
\section{Evaluation Results}
\label{sec:result}

In this section, we present the results of our experimental evaluations and address the research questions posed in Section~\ref{sec:rq}. The experiments were run in parallel on several machines each with six cores CPU at 2.9GHz and 8GB RAM for two months ($24 \times 7$). All settings discussed in Section~\ref{sec:exp} are used unless otherwise stated. 


\begin{figure}[b!]
\centering
\includegraphics[width=0.87\columnwidth]{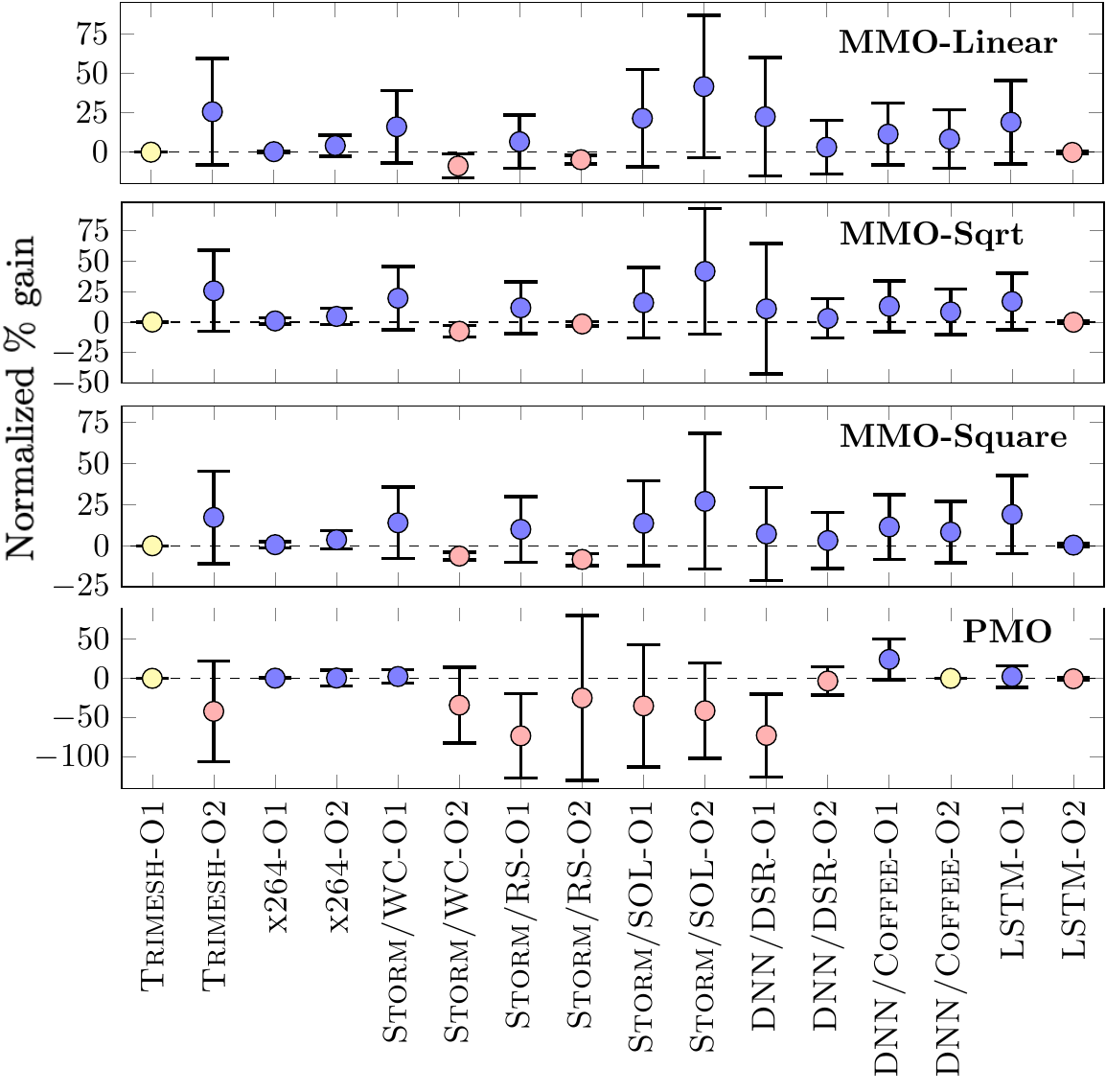}
\caption{Average and standard deviations on \% gain of MMO and PMO over the best single-objective counterpart for 30 runs. The \setlength{\fboxsep}{1.5pt}\colorbox{steel!30}{blue}, \setlength{\fboxsep}{1.5pt}\colorbox{red!10}{pink}, and \setlength{\fboxsep}{1.5pt}\colorbox{yellow!30}{yellow} denote positive, negative, and zero average gain, respectively.}
\label{fig:gain}
\end{figure}


\subsection{RQ1: Effectiveness}
\label{sec:rq1}

\subsubsection{Method}

To answer \textbf{RQ1}, we examine all the eight software systems/environments with two performance objectives each, giving us 16 cases of study. In each case, we compare the best single-objective counterpart\footnote{We identified the best one among RS, SHC-r, SOGA, and SA based on the best rank from the Scott-Knott test~\cite{scott1974cluster} over 30 runs for stronger statistical power. If multiple optimizers share the best rank, we picked the one with the best average result. Note that the best single-objective counterpart may differ case by case.} with all instances of MMO and PMO. To set the weight for each MMO instance in a case, we firstly conduct preliminary runs under 10\% of the search budget and population size (one run each value). The weight with the best target performance objective is then used in the full-scale experiments (if more than one weight is the best, we chose one randomly). For all pair-wise comparisons, both Wilcoxon sign-rank test and $\hat{A}_{12}$ are used. 




\input{tables/quality}

\subsubsection{Results}

From Figure~\ref{fig:gain}, we see that, on average, the MMO model achieves reasonably positive gains for at least 12 cases across the instances. This is a sign that the MMO model suffers less on the local optima issue than its best single-objective counterpart. Moreover, 
it achieves improvements for more than an average of 30\% in some cases, e.g., \textsc{Storm/SOL-O2}, as in those cases the distance between local optima can be large. 
Yet, we observe no obvious difference among the MMO instances. 
PMO, albeit leads to acceptable results for some cases, often performs worse than the best single-objective counterpart as the gains are generally similar or negative (11 out of 16 cases). 
This can be attributed to the fact that it wastes a significant amount of resources on optimizing the auxiliary performance objective. 
Interestingly, for \textsc{Trimesh-O1}, all the models have the same results as the best single-objective counterpart. 
Although rare, this is a possible case where the landscape of the target performance objective is simpler (e.g., fewer local optima); 
hence all the models/optimizers can find the globally optimal configuration.



Table~\ref{tb:gain} shows the results of the statistical tests, in which we see that similar to the gains, the MMO model wins a larger majority in general, in which most of them are statistically significant ($p<0.05$) with non-trivial effect size ($\hat{A}_{12} \geq 0.56$). Again, the PMO performs the worst with no wins on 12 out of 16 cases. 




To provide a detailed understanding, Figure~\ref{fig:detail} shows all the explored configurations for \textsc{Trimesh} with latency as the target performance objective. Clearly, we see that the result confirms our theory: the single-objective counterparts do explore some good ranges of configurations, but they remain mostly trapped in a large region of local optima. The PMO performs the worst with fewer points in the projected area because it over-empathizes on optimizing the auxiliary performance objective, which negatively affects the target performance objective. Our MMO model, in contrast, escapes from local optima by exploring an even larger area while keeping the tendency towards better target performance objective, which is precisely our \textbf{Goals 1} and \textbf{2} from Section~\ref{sec:method}. Therefore, we say:

\begin{figure}[t!]
\centering
\includegraphics[width=0.7\columnwidth]{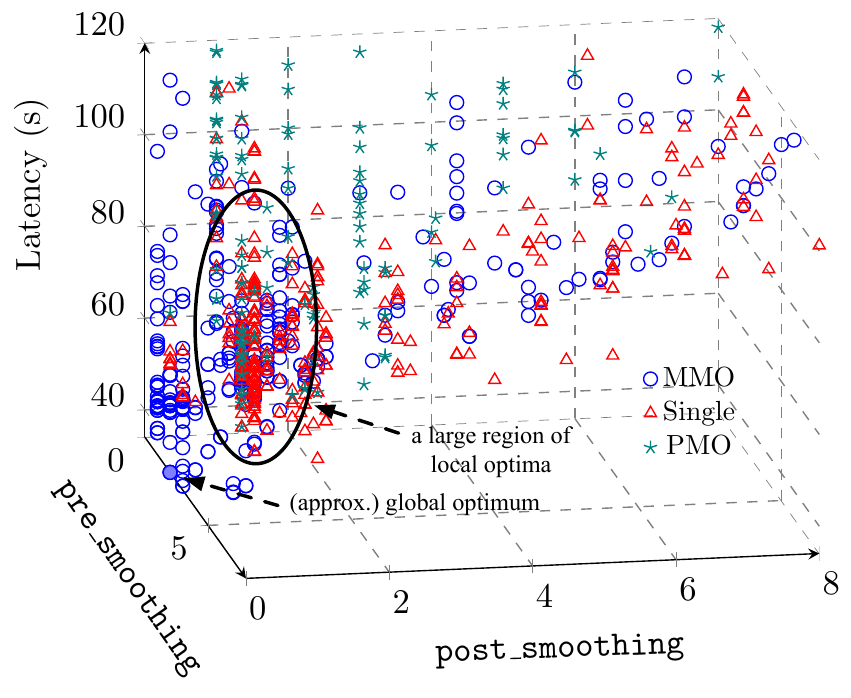}
\caption{A projected landscape of \textsc{Trimesh} explored by MMO (since all instances perform similarly, here we use MMO-Linear as an example), PMO, and the best single-objective counterpart. Each point is a configuration measured in the run, regardless whether it is preserved or not.}
\label{fig:detail}
\end{figure}

\begin{quotebox}
   \noindent
   \textit{\textbf{RQ1:} The MMO model, regardless of its instance, is effective in overcoming local optima, providing considerably better results than the best single-objective counterpart in general (up to 42\% mean gain). It also significantly outperforms the PMO model. The MMO instances do not differ much though.}
\end{quotebox}

\subsection{RQ2: Resource Efficiency}
\label{sec:rq2}

\subsubsection{Method}

To investigate \textbf{RQ2}, for each case, we use a baseline, $b$, taken as the smallest number of measurements that the best single-objective counterpart consumes to achieve its best average result over 30 runs. We then record the smallest amount of budget consumed by the MMO and PMO to achieve the same (or better) target performance objective on average, denoted as $m$. The ratios, i.e., $r={m \over b} \times 100\%$, are reported, implying that if the MMO instances are resource-efficient, then we would expect $r \leq 100\%$. Since in our context the resource is the number of measurements, it reflects the tuning time and computation required by a model. Again, as for \textbf{RQ1}, only the best weight for each MMO instance identified from the preliminary runs is examined in a case.



\subsubsection{Results}


As can be seen from Figure~\ref{fig:resource}, despite a small number of cases 
where the MMO model cannot reach the performance level as achieved by the best single-objective counterpart (the divided bars, denoted as $r \gg 100\%$),
most commonly it uses less number of measurements than, or at least identical to, the baseline to find the same or better results, e.g., it can be as significantly low as 24\%. In particular, the MMO instances have 10-13 cases of $r < 100\%$; 1-3 cases of $r = 100\%$; and 2-3 cases of $r \gg 100\%$. This indicates that the MMO model overcomes local optima better and more efficiently --- a key attraction to software configuration tuning due to its expensive measurements. In contrast, the PMO exhibits the worst resource efficiency, as it has 3 cases of $r < 100\%$, together with 1 and 2 cases of $r = 100\%$ and $r > 100\%$, respectively, while the remaining 10 cases of $r \gg 100\%$. This is a clear sign that PMO is generally resource-hungry as discussed in Section~\ref{sec:method}.

Notably, the resource saving of MMO is more significant on systems with larger search space, e.g., \textsc{Keras-DNN} and \textsc{Trimesh}. 
This is because that the larger the search space, 
the more the resources are required for the single-objective model to find good configurations. 
In contrast, our model MMO, 
which is designed to keep a set of diverse high-quality configurations during the tuning,
needs less effort to find better ones. 
As a result, we conclude that:

\begin{figure}[t!]
\centering
\includegraphics[width=\columnwidth]{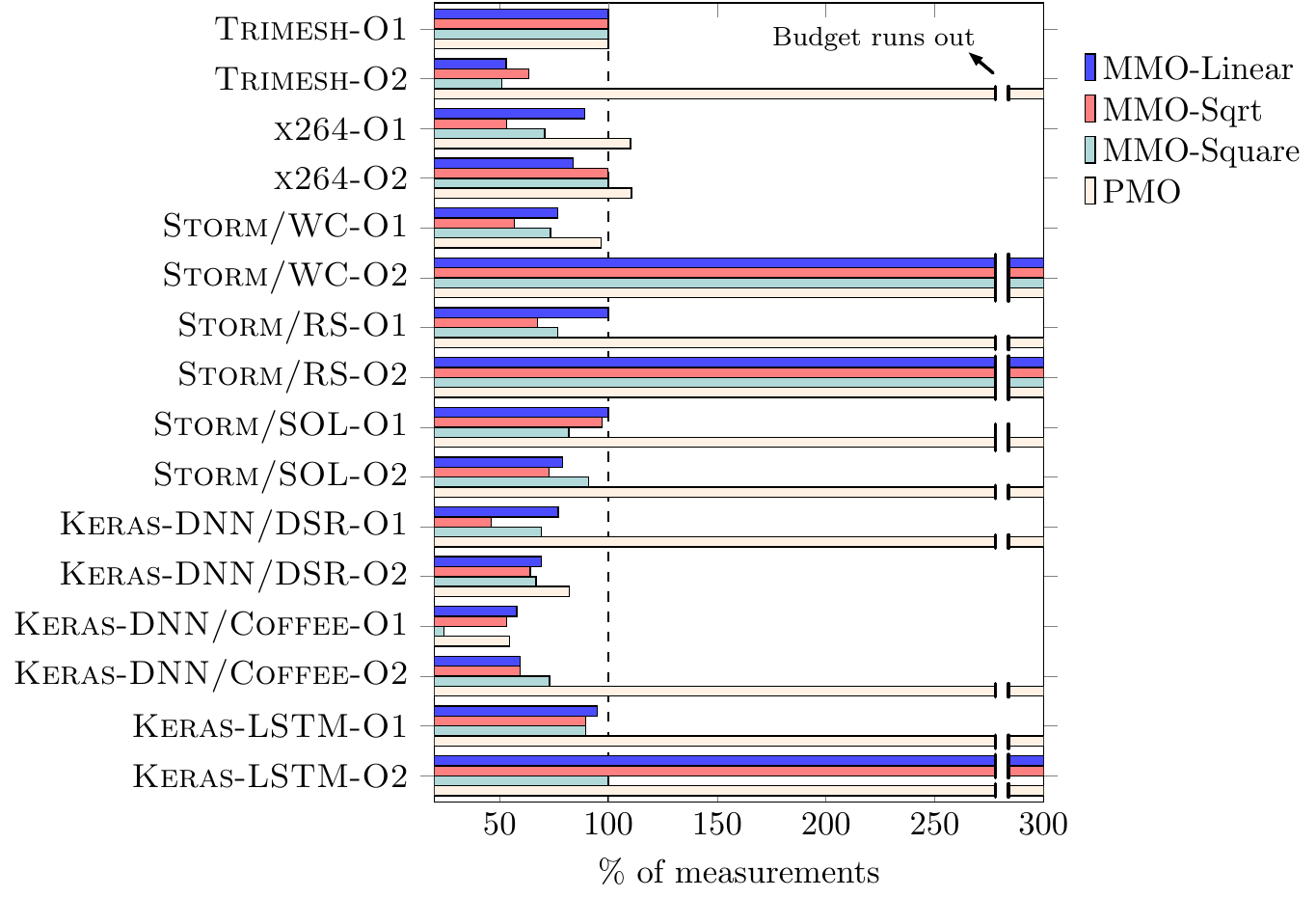}
\caption{\% of measurements ($r$) for the MMO and PMO models to converge to the best (average) performance objective by the best single-objective counterpart over 30 runs, using its budget consumption as the baseline (the dashed line). The divided bars denote no convergence when the total search budget runs out, i.e., $r \gg 100\%$.}
\label{fig:resource}

\end{figure}

\begin{quotebox}
   \noindent
   \textit{\textbf{RQ2:} The MMO model is resource-efficient, consuming generally fewer measurements than the best single-objective counterpart to reach the same or better results (as low as 24\% of it). The PMO, in contrast, is much more resource-hungry.}
\end{quotebox}

\subsection{RQ3: Sensitivity to Weight}
\label{sec:rq3}

\subsubsection{Method}

To address \textbf{RQ3}, we check how do the MMO instances perform compared with the best single-objective counterpart under different weight settings for the full-scale experiments. Hence, for each instance, there are seven settings and 16 systems/environments, leading to 112 cases. In each of these cases, we conduct a pair-wise comparison using the $\hat{A}_{12}$ and Wilcoxon sign-rank test.

\subsubsection{Results}

The results are shown in Table~\ref{tb:sensitivity}, in which we see that the MMO model, regardless to its instance, may indeed be sensitive to the weight as it could win or lose (with different $\hat{A}_{12}$ values and statistical results) depending on different settings. Although the best weight can be different for specific cases, we do observe a general pattern: according to the last row, setting the weight as an edge value like 0.01, 0.1, 0.9, or 10 tends to be the best among others in general. This is clearer for MMO-Linear and {MMO-Sqrt}, while {MMO-Square} prefers 0.01 more. We also note that the best weights identified from the preliminary runs are generally consistent with those best ones under the full-scale experiments.



In particular, we see that all the MMO instances can be more beneficial (more weight values, if not all, can lead to significantly better results) in some cases of the complex systems (e.g., \textsc{x264} and \textsc{Keras-DNN}) than others with smaller search space and dimension of options (e.g., \textsc{Storm}). 
This could be due to the fact that for more challenging systems,
the advantage of our model over the single-objective counterpart is clearer, 
thus it is easier to have a better result over different weight settings. 
In summary, we state that:



\begin{quotebox}
   \noindent
   \textit{\textbf{RQ3:} The MMO model is sensitive to the weight, but there exist a common pattern such that some extreme weight, e.g., 0.01, 0.1, 0.9 or 10, is often the best value. }
\end{quotebox}






%% file: tables/quality.tex
\begin{table}[b!]
\caption{$\hat{A}_{12}$ and $p$ values on comparing multi-objectivization (MMO and PMO model) against the best single-objective counterpart over 30 runs.}
\label{tb:gain}
\setlength{\tabcolsep}{0.8mm}
\centering
\footnotesize
\begin{tabular}{lcccc}\toprule

\textbf{Software System}&\textbf{{MMO-Linear}}&\textbf{{MMO-Sqrt}}&\textbf{{MMO-Square}}&\textbf{PMO}\\

\midrule

\textsc{Trimesh-O1}&.50 ($<$.001)&.50 ($<$.001)&.50 ($<$.001)&.50 ($<$.001)\\ 

\textsc{Trimesh-O2}&\cellcolor{steel!30}\textbf{.88 ($<$.001)}&\cellcolor{steel!30}\textbf{.93 ($<$.001)}&\cellcolor{steel!30}\textbf{.88 ($<$.001)}&\cellcolor{red!10}\textbf{.25 ($=$.002)}\\ 

\hline 

\textsc{x264-O1}&\cellcolor{steel!30}\textbf{.80 ($<$.001)}&\cellcolor{steel!30}\textbf{.73 ($<$.001)}&\cellcolor{steel!30}\textbf{.97 ($<$.001)}&\cellcolor{steel!30}\textbf{.85 ($<$.001)}\\ 

\textsc{x264-O2}&\cellcolor{steel!30}\textbf{.78 ($<$.001)}&\cellcolor{steel!30}\textbf{.82 ($<$.001)}&\cellcolor{steel!30}\textbf{.77 ($<$.001)}&\cellcolor{red!10}.40 ($=$.918)\\ 

\hline 

\textsc{Storm/WC-O1}&\cellcolor{steel!30}\textbf{.67 ($<$.001)}&\cellcolor{steel!30}\textbf{.68 ($<$.001)}&\cellcolor{steel!30}\textbf{.65 ($<$.001)}&\cellcolor{steel!30}.53 ($<$.001)\\ 

\textsc{Storm/WC-O2}&\cellcolor{red!10}\textbf{.03 ($<$.001)}&\cellcolor{red!10}\textbf{.02 ($<$.001)}&\cellcolor{red!10}\textbf{.00 ($<$.001)}&\cellcolor{red!10}.33 ($=$.636)\\ 

\hline 

\textsc{Storm/RS-O1}&\cellcolor{steel!30}\textbf{.57 ($<$.001)}&\cellcolor{steel!30}\textbf{.62 ($<$.001)}&\cellcolor{steel!30}\textbf{.60 ($<$.001)}&\cellcolor{red!10}\textbf{.10 ($<$.001)}\\ 

\textsc{Storm/RS-O2}&\cellcolor{red!10}\textbf{.00 ($<$.001)}&\cellcolor{red!10}\textbf{.17 ($<$.001)}&\cellcolor{red!10}\textbf{.02 ($<$.001)}&\cellcolor{red!10}.47 ($<$.001)\\ 

\hline 

\textsc{Storm/SOL-O1}&\cellcolor{steel!30}\textbf{.67 ($<$.001)}&\cellcolor{steel!30}\textbf{.62 ($<$.001)}&\cellcolor{steel!30}\textbf{.62 ($<$.001)}&\cellcolor{red!10}.40 ($=$.334)\\ 

\textsc{Storm/SOL-O2}&\cellcolor{steel!30}\textbf{.72 ($<$.001)}&\cellcolor{steel!30}\textbf{.72 ($<$.001)}&\cellcolor{steel!30}\textbf{.65 ($<$.001)}&\cellcolor{red!10}.28 ($=$.100)\\ 

\hline 

\textsc{Keras-DNN/DSR-O1}&\cellcolor{steel!30}\textbf{.67 ($=$.001)}&\cellcolor{steel!30}\textbf{.62 ($=$.031)}&\cellcolor{steel!30}.53 ($=$.008)&\cellcolor{red!10}\textbf{.05 ($<$.001)}\\ 

\textsc{Keras-DNN/DSR-O2}&\cellcolor{steel!30}.52 ($<$.001)&\cellcolor{steel!30}.52 ($<$.001)&\cellcolor{steel!30}.52 ($<$.001)&\cellcolor{red!10}.48 ($<$.001)\\ 

\hline 

\textsc{Keras-DNN/Coffee-O1}&\cellcolor{steel!30}\textbf{.67 ($<$.001)}&\cellcolor{steel!30}\textbf{.68 ($<$.001)}&\cellcolor{steel!30}\textbf{.67 ($<$.001)}&\cellcolor{steel!30}\textbf{.73 ($<$.001)}\\ 

\textsc{Keras-DNN/Coffee-O2}&\cellcolor{steel!30}\textbf{.58 ($<$.001)}&\cellcolor{steel!30}\textbf{.58 ($<$.001)}&\cellcolor{steel!30}\textbf{.58 ($<$.001)}&.50 ($<$.001)\\ 

\hline 

\textsc{Keras-LSTM-O1}&\cellcolor{steel!30}\textbf{.68 ($<$.001)}&\cellcolor{steel!30}\textbf{.70 ($<$.001)}&\cellcolor{steel!30}\textbf{.72 ($<$.001)}&\cellcolor{steel!30}.53 ($<$.001)\\ 

\textsc{Keras-LSTM-O2}&\cellcolor{red!10}.48 ($=$.002)&\cellcolor{red!10}.48 ($=$.002)&\cellcolor{steel!30}\textbf{.58 ($<$.001)}&\cellcolor{red!10}.42 ($=$.056)\\

\bottomrule
\end{tabular}
\centering
 \begin{tablenotes}
    \footnotesize
    \item The $p$ values are shown in the bracket. $\hat{A}_{12}>0.5$ means the MMO (or PMO) is better (in {\setlength{\fboxsep}{1.5pt}\colorbox{steel!30}{blue}}); $\hat{A}_{12}<0.5$ denotes the best single-objective counterpart is better (in {\setlength{\fboxsep}{1.5pt}\colorbox{red!10}{pink}}); $\hat{A}_{12}=0.5$ means a tie. The comparisons, for which there is a $p<$0.05 and $\hat{A}_{12} \leq 0.44$ or $\hat{A}_{12} \geq 0.56$, are highlighted in \textbf{bold}.
    
    \end{tablenotes}
\end{table}

%% file: discussion.tex
\section{Discussion}
\label{sec:discussion}

\subsection{Why Software Configuration Tuning?}

A natural question to ask is why our MMO model is specific for software configuration tuning rather than as a general ``search-based'' solution for all SBSE problems. The answer is three-fold. 


    Firstly, we took three important properties of software configuration tuning into account when designing the MMO model: (1) the high degree of sparsity in software systems exacerbates the issue of the search being trapped in local optima (\textbf{Property 1})~\cite{nair2018finding,DBLP:conf/mascots/JamshidiC16}. (2) The measurement is expensive; 
    thus efficiently escaping the local optima with less resources is desirable (\textbf{Property 2}). (3) The correlations between performance attributes are uncertain, i.e., extremely well or poor auxiliary performance objective may both lead to similarly good target performance objective (\textbf{Property 3}). 
    \input{tables/sensitivity}

    Secondly, the configurable software systems provide a well-fitting avenue for multi-objectivization, as it is common that they inherently come with at least two performance attributes, e.g., latency and throughput, that can be directly used in the multi-objectivization. Some other SBSE problems, in contrast, may not have a readily available attribute(s) that can serve as the auxiliary objective. For example, in the code refactoring problem, the robustness of the code (as an auxiliary objective) is not a straightforward and widely known metric that can be easily quantified~\cite{DBLP:conf/ssbse/MkaouerKBC14}. 
    
    
    Finally, how the configurations can affect the performance attributes is often a black-box. In contrast, in many other SBSE problems, the objective function can be specifically designed based on some shared domain properties between scenarios. For example, in crash reproduction problem~\cite{derakhshanfar2020good}, it is possible to engineer a new auxiliary objective to check how widely a test case covers the code based on some common code structures for a software system (e.g., function access levels), such that it is strongly conflicting with the target objective (i.e., distance to the particular line(s)-of-code that reproduces the crash). However, it is difficult, if not impossible, for the configuration options in software configuration tuning to achieve the same. Our MMO model explicitly considers such a black-box nature of software configuration tuning, as we make no assumption about the performance attributes and their correlations.











\begin{figure}[!t]
\centering
\includegraphics[width=\columnwidth]{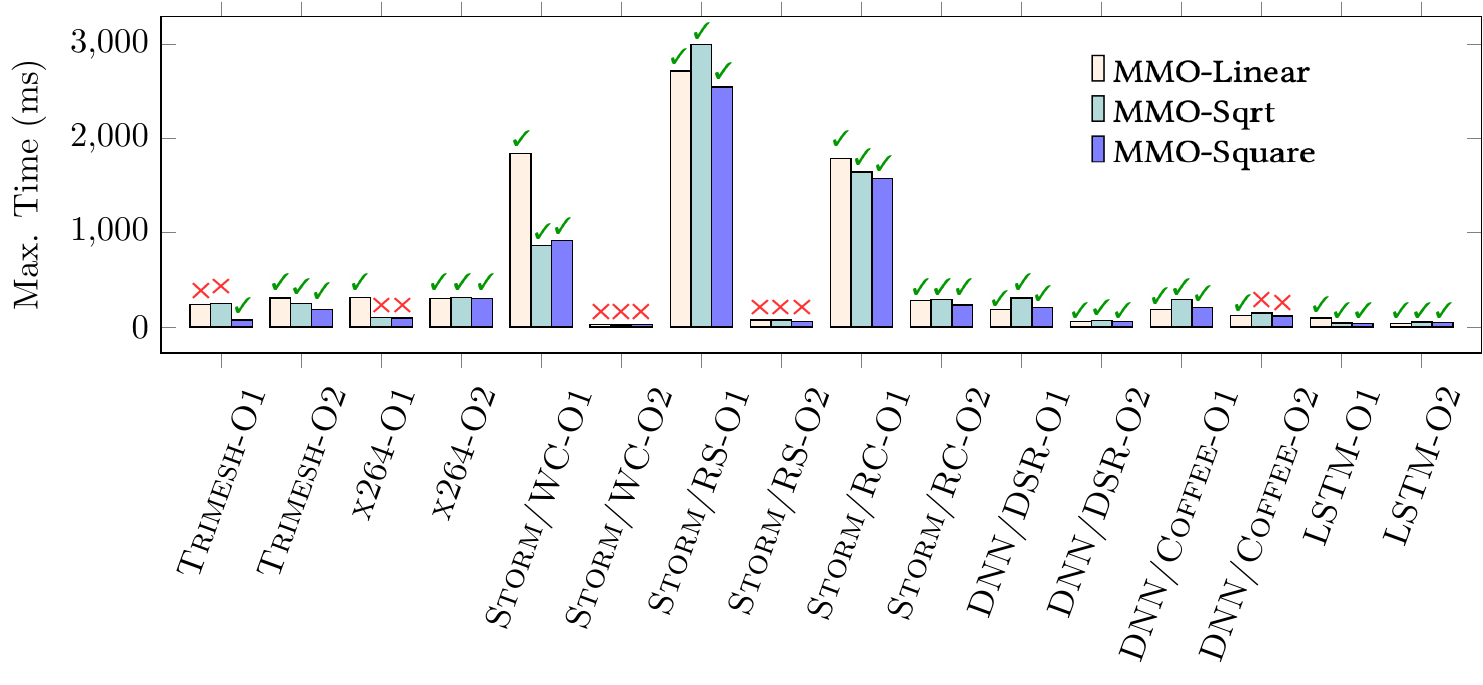}
\caption{The maximum running time for a full-scale run when using previously measured data to identify the best weight (over all seven weight values and 30 runs). \textcolor{green!60!black}{\ding{51}} denotes the best weight concluded using data is identical to that obtained from profiling the system; \textcolor{red!80}{\ding{53}} means otherwise.}
\label{fig:data-time}
\end{figure}

\subsection{How to Use MMO in Practice?}
\label{sec:weight-discussion}

Here, we elaborate on the guidelines for using MMO in practice. 

\subsubsection{Choosing the MMO instance}

Sections~\ref{sec:rq1} and~\ref{sec:rq2} reveal that 
the MMO instances perform similarly for software configuration tuning --- 
all better than the best single-objective counterpart and PMO in general.
It is, therefore, safe to choose any of them. In general, we suggest to use MMO-Linear by default as it is the simplest form among the others.

\subsubsection{Setting the weight}

We recommend two alternative methods to identify the best weights during preliminary runs: physically profiled method and data-driven method.

To set a good weight, one can profile the actual system with a reduced budget (e.g., 10\%), as what we have done in this work. In fact, the findings from Section~\ref{sec:rq3} has provided useful insights to simplify the process: albeit the difficulty of setting weight varies depending on the case, we observed that the edge weight value, e.g., 0.01, 0.1, 0.9 or 10, is generally a reliable setting\footnote{We have additionally examined values $<0.01$ or $>10$ in our experiments, but the results make no statistically significant improvements across the cases.}. Further, there are also cases where nearly all weights we examined are highly effective, such as \textsc{x264-O1} and \textsc{Keras-DNN/Coffee-O2}. Therefore, we suggest trying at least the above values in the preliminary runs. 


When previously measured data is available, the data-driven method for identifying the weight becomes possible. We have found that, for all the MMO instances and cases under the full-scale experiments, the best weight value concluded based on the data is the same as that identified by measuring the system, but the former can terminate several orders of magnitude faster. As shown in Figure~\ref{fig:data-time}, when examining the weight using all data collected from the previous experiments, we see that the resulted best weights are generally consistent with those identified by physically profiling the systems (as in Section~\ref{sec:rq3}). For the few cases where there is an inconsistency, all the weights in fact perform rather similar (e.g., \textsc{x264-O1}), therefore it is safe even if the actual best one has not been chosen. More importantly, the maximum running time is negligible when using data --- it is merely 3 seconds or less.






\subsection{Threats to Validity}


Threats to \textbf{internal validity} can be related to the search budget. To tackle this, we have used two-hour budgets as suggested in prior work~\cite{DBLP:conf/mascots/JamshidiC16}. The parameter settings follow what has been used from the literature or tuned through preliminary runs. To mitigate bias, we repeated 30 experiment runs under each case.

The metrics and evaluation used may pose threats to \textbf{construct validity}. Since there is only a single performance concern, 
we conduct the comparison based on the gains on the target performance objectives over the best single-objective optimizer, 
together with the resources (number of measurements) required to converge to the same result. Both of these are common metrics in software configuration tuning~\cite{nair2018finding}. To verify statistical significance and effect size, we use Wilcoxon sign-rank test and $\hat{A}_{12}$ to examine the results. 

Threats to \textbf{external validity} can be raised from the subjects studied. We mitigated this by using eight systems/environments that are of different scales and performance attributes. We also compared the MMO model with four state-of-the-art single-objective counterparts for software configuration tuning. Nonetheless, we agree that using more systems and optimizers may prove fruitful.

%% file: tables/sensitivity.tex
\begin{table*}[t!]
\caption{Sensitivity analysis on different weights in the MMO model (full-scale experiments). The cells report the $\hat{A}_{12}$ values and whether $p<0.05$ on comparing a MMO instance and the best single-objective counterpart over 30 runs. The last row counts how many times a weight value is the best in a case based on Scott-Knott rank (primary) and the average result (secondary).}
\label{tb:sensitivity}
\setlength{\tabcolsep}{1.4mm}
\centering
\footnotesize
\begin{tabular}{lccccccc||ccccccc||ccccccc}\toprule

\multirow{2}{*}{\textbf{Software System}}&\multicolumn{7}{c||}{\textbf{{MMO-Linear}}}&\multicolumn{7}{c||}{\textbf{{MMO-Sqrt}}}&\multicolumn{7}{c}{\textbf{{MMO-Square}}}\\ \cmidrule{2-22}

&\textbf{0.01}&\textbf{0.1}&\textbf{0.3}&\textbf{0.5}&\textbf{0.7}&\textbf{0.9}&\textbf{10}&\textbf{0.01}&\textbf{0.1}&\textbf{0.3}&\textbf{0.5}&\textbf{0.7}&\textbf{0.9}&\textbf{10}&\textbf{0.01}&\textbf{0.1}&\textbf{0.3}&\textbf{0.5}&\textbf{0.7}&\textbf{0.9}&\textbf{10}\\

\midrule

\textsc{Trimesh-O1}&.50$^\dagger$&.50$^\dagger$&.50$^\dagger$&.50$^\dagger$&.50$^\dagger$&.50$^\dagger$&.50$^\dagger$&.50$^\dagger$&.50$^\dagger$&.50$^\dagger$&.50$^\dagger$&.50$^\dagger$&.50$^\dagger$&.50$^\dagger$&.50$^\dagger$&.50$^\dagger$&.50$^\dagger$&.50$^\dagger$&.50$^\dagger$&.50$^\dagger$&.50$^\dagger$\\ 
\textsc{Trimesh-O2}&\cellcolor{steel!30}\textbf{.75$^\dagger$}&\cellcolor{steel!30}\textbf{.88$^\dagger$}&\cellcolor{red!10}\textbf{.07$^\dagger$}&\cellcolor{red!10}\textbf{.00$^\dagger$}&\cellcolor{red!10}\textbf{.00$^\dagger$}&\cellcolor{red!10}\textbf{.00$^\dagger$}&\cellcolor{red!10}\textbf{.00$^\dagger$}&\cellcolor{red!10}.35&\cellcolor{steel!30}\textbf{.93$^\dagger$}&\cellcolor{red!10}\textbf{.20$^\dagger$}&\cellcolor{red!10}\textbf{.00$^\dagger$}&\cellcolor{red!10}\textbf{.05$^\dagger$}&\cellcolor{red!10}\textbf{.08$^\dagger$}&\cellcolor{red!10}\textbf{.05$^\dagger$}&\cellcolor{steel!30}\textbf{.88$^\dagger$}&\cellcolor{red!10}\textbf{.00$^\dagger$}&\cellcolor{red!10}\textbf{.02$^\dagger$}&\cellcolor{red!10}\textbf{.02$^\dagger$}&\cellcolor{red!10}\textbf{.00$^\dagger$}&\cellcolor{red!10}\textbf{.05$^\dagger$}&\cellcolor{red!10}\textbf{.02$^\dagger$}\\ \hline
\textsc{x264-O1}&\cellcolor{steel!30}\textbf{.83$^\dagger$}&\cellcolor{steel!30}\textbf{.90$^\dagger$}&\cellcolor{steel!30}\textbf{.82$^\dagger$}&\cellcolor{steel!30}\textbf{.80$^\dagger$}&\cellcolor{steel!30}\textbf{.87$^\dagger$}&\cellcolor{steel!30}\textbf{.88$^\dagger$}&\cellcolor{steel!30}\textbf{.80$^\dagger$}&\cellcolor{steel!30}\textbf{.95$^\dagger$}&\cellcolor{steel!30}\textbf{.93$^\dagger$}&\cellcolor{steel!30}\textbf{.90$^\dagger$}&\cellcolor{steel!30}\textbf{.88$^\dagger$}&\cellcolor{steel!30}\textbf{.85$^\dagger$}&\cellcolor{steel!30}\textbf{.82$^\dagger$}&\cellcolor{steel!30}\textbf{.73$^\dagger$}&\cellcolor{steel!30}\textbf{.88$^\dagger$}&\cellcolor{steel!30}\textbf{.87$^\dagger$}&\cellcolor{steel!30}\textbf{.80$^\dagger$}&\cellcolor{steel!30}\textbf{.83$^\dagger$}&\cellcolor{steel!30}\textbf{.97$^\dagger$}&\cellcolor{steel!30}\textbf{.88$^\dagger$}&\cellcolor{steel!30}\textbf{.72$^\dagger$}\\ 
\textsc{x264-O2}&\cellcolor{steel!30}\textbf{.58$^\dagger$}&\cellcolor{steel!30}\textbf{.60$^\dagger$}&\cellcolor{red!10}.38&\cellcolor{steel!30}\textbf{.78$^\dagger$}&.50&\cellcolor{red!10}.45&\cellcolor{red!10}.48&\cellcolor{red!10}.47&\cellcolor{red!10}.42&\cellcolor{red!10}\textbf{.27$^\dagger$}&\cellcolor{red!10}.45&\cellcolor{red!10}.40&\cellcolor{steel!30}\textbf{.58$^\dagger$}&\cellcolor{steel!30}\textbf{.82$^\dagger$}&\cellcolor{steel!30}\textbf{.77$^\dagger$}&\cellcolor{red!10}.47&.50&\cellcolor{steel!30}.53$^\dagger$&\cellcolor{red!10}.40&\cellcolor{red!10}.48&\cellcolor{red!10}.43\\ \hline
\textsc{Storm/WC-O1}&\cellcolor{red!10}.39&\cellcolor{steel!30}\textbf{.62$^\dagger$}&\cellcolor{red!10}.45$^\dagger$&\cellcolor{red!10}\textbf{.43$^\dagger$}&\cellcolor{steel!30}\textbf{.67$^\dagger$}&\cellcolor{steel!30}\textbf{.63$^\dagger$}&\cellcolor{red!10}.33&\cellcolor{steel!30}\textbf{.58$^\dagger$}&\cellcolor{steel!30}.52$^\dagger$&\cellcolor{steel!30}.52$^\dagger$&\cellcolor{red!10}\textbf{.43$^\dagger$}&\cellcolor{steel!30}.53$^\dagger$&\cellcolor{steel!30}\textbf{.68$^\dagger$}&\cellcolor{red!10}.38&\cellcolor{steel!30}\textbf{.58$^\dagger$}&\cellcolor{steel!30}\textbf{.58$^\dagger$}&\cellcolor{steel!30}\textbf{.60$^\dagger$}&\cellcolor{steel!30}\textbf{.65$^\dagger$}&\cellcolor{steel!30}.53$^\dagger$&\cellcolor{steel!30}.52$^\dagger$&\cellcolor{red!10}.30\\ 
\textsc{Storm/WC-O2}&\cellcolor{red!10}\textbf{.03$^\dagger$}&\cellcolor{red!10}\textbf{.00$^\dagger$}&\cellcolor{red!10}\textbf{.00$^\dagger$}&\cellcolor{red!10}\textbf{.00$^\dagger$}&\cellcolor{red!10}\textbf{.00$^\dagger$}&\cellcolor{red!10}\textbf{.00$^\dagger$}&\cellcolor{red!10}\textbf{.02$^\dagger$}&\cellcolor{red!10}\textbf{.02$^\dagger$}&\cellcolor{red!10}\textbf{.00$^\dagger$}&\cellcolor{red!10}\textbf{.02$^\dagger$}&\cellcolor{red!10}\textbf{.00$^\dagger$}&\cellcolor{red!10}\textbf{.00$^\dagger$}&\cellcolor{red!10}\textbf{.00$^\dagger$}&\cellcolor{red!10}\textbf{.00$^\dagger$}&\cellcolor{red!10}\textbf{.00$^\dagger$}&\cellcolor{red!10}\textbf{.03$^\dagger$}&\cellcolor{red!10}\textbf{.02$^\dagger$}&\cellcolor{red!10}\textbf{.00$^\dagger$}&\cellcolor{red!10}\textbf{.02$^\dagger$}&\cellcolor{red!10}\textbf{.00$^\dagger$}&\cellcolor{red!10}\textbf{.02$^\dagger$}\\ \hline
\textsc{Storm/RS-O1}&\cellcolor{steel!30}.55$^\dagger$&\cellcolor{steel!30}.52$^\dagger$&\cellcolor{red!10}.42&\cellcolor{red!10}.48$^\dagger$&\cellcolor{steel!30}\textbf{.57$^\dagger$}&\cellcolor{steel!30}\textbf{.57$^\dagger$}&\cellcolor{red!10}\textbf{.10$^\dagger$}&\cellcolor{steel!30}.55$^\dagger$&\cellcolor{steel!30}.55$^\dagger$&\cellcolor{steel!30}.53$^\dagger$&\cellcolor{steel!30}.52$^\dagger$&\cellcolor{steel!30}.53$^\dagger$&\cellcolor{steel!30}\textbf{.62$^\dagger$}&\cellcolor{red!10}\textbf{.18$^\dagger$}&\cellcolor{steel!30}\textbf{.57$^\dagger$}&\cellcolor{steel!30}\textbf{.60$^\dagger$}&\cellcolor{steel!30}\textbf{.57$^\dagger$}&.50$^\dagger$&\cellcolor{steel!30}.53$^\dagger$&\cellcolor{steel!30}.52$^\dagger$&\cellcolor{red!10}\textbf{.22$^\dagger$}\\ 
\textsc{Storm/RS-O2}&\cellcolor{red!10}\textbf{.00$^\dagger$}&\cellcolor{red!10}\textbf{.00$^\dagger$}&\cellcolor{red!10}\textbf{.00$^\dagger$}&\cellcolor{red!10}\textbf{.00$^\dagger$}&\cellcolor{red!10}\textbf{.02$^\dagger$}&\cellcolor{red!10}\textbf{.00$^\dagger$}&\cellcolor{red!10}\textbf{.00$^\dagger$}&\cellcolor{red!10}\textbf{.17$^\dagger$}&\cellcolor{red!10}\textbf{.00$^\dagger$}&\cellcolor{red!10}\textbf{.00$^\dagger$}&\cellcolor{red!10}\textbf{.00$^\dagger$}&\cellcolor{red!10}\textbf{.00$^\dagger$}&\cellcolor{red!10}\textbf{.00$^\dagger$}&\cellcolor{red!10}\textbf{.00$^\dagger$}&\cellcolor{red!10}\textbf{.02$^\dagger$}&\cellcolor{red!10}\textbf{.00$^\dagger$}&\cellcolor{red!10}\textbf{.00$^\dagger$}&\cellcolor{red!10}\textbf{.00$^\dagger$}&\cellcolor{red!10}\textbf{.00$^\dagger$}&\cellcolor{red!10}\textbf{.00$^\dagger$}&\cellcolor{red!10}\textbf{.00$^\dagger$}\\ \hline
\textsc{Storm/SOL-O1}&\cellcolor{red!10}\textbf{.43$^\dagger$}&\cellcolor{red!10}.45$^\dagger$&\cellcolor{red!10}.47$^\dagger$&\cellcolor{steel!30}\textbf{.57$^\dagger$}&\cellcolor{steel!30}\textbf{.67$^\dagger$}&\cellcolor{steel!30}\textbf{.58$^\dagger$}&\cellcolor{red!10}\textbf{.18$^\dagger$}&\cellcolor{red!10}.45$^\dagger$&\cellcolor{red!10}.47$^\dagger$&\cellcolor{red!10}.40&\cellcolor{steel!30}.52$^\dagger$&\cellcolor{steel!30}\textbf{.62$^\dagger$}&\cellcolor{steel!30}.55$^\dagger$&\cellcolor{red!10}\textbf{.18$^\dagger$}&\cellcolor{red!10}.38&\cellcolor{red!10}.48$^\dagger$&.50$^\dagger$&\cellcolor{red!10}\textbf{.43$^\dagger$}&\cellcolor{red!10}.45$^\dagger$&\cellcolor{steel!30}\textbf{.62$^\dagger$}&\cellcolor{steel!30}.52$^\dagger$\\ 
\textsc{Storm/SOL-O2}&\cellcolor{steel!30}\textbf{.72$^\dagger$}&\cellcolor{steel!30}.53$^\dagger$&\cellcolor{red!10}.43&\cellcolor{steel!30}.53$^\dagger$&\cellcolor{red!10}.42&\cellcolor{red!10}.30&\cellcolor{red!10}\textbf{.17$^\dagger$}&\cellcolor{steel!30}\textbf{.72$^\dagger$}&\cellcolor{steel!30}.55$^\dagger$&\cellcolor{steel!30}.53$^\dagger$&\cellcolor{red!10}.38&\cellcolor{red!10}.38&\cellcolor{red!10}\textbf{.25$^\dagger$}&\cellcolor{red!10}\textbf{.17$^\dagger$}&\cellcolor{steel!30}\textbf{.65$^\dagger$}&\cellcolor{steel!30}.55$^\dagger$&\cellcolor{red!10}.38&\cellcolor{red!10}.45&\cellcolor{red!10}.35&\cellcolor{red!10}.28&\cellcolor{red!10}.33\\ \hline
\textsc{Keras-DNN/DSR-O1}&\cellcolor{red!10}\textbf{.22$^\dagger$}&\cellcolor{red!10}.33&\cellcolor{red!10}.45&\cellcolor{steel!30}\textbf{.57$^\dagger$}&\cellcolor{red!10}.47&\cellcolor{red!10}\textbf{.20$^\dagger$}&\cellcolor{steel!30}\textbf{.67$^\dagger$}&\cellcolor{red!10}.30&\cellcolor{red!10}\textbf{.25$^\dagger$}&\cellcolor{red!10}\textbf{.28$^\dagger$}&\cellcolor{red!10}.33&\cellcolor{red!10}.42&\cellcolor{red!10}.47&\cellcolor{steel!30}\textbf{.62$^\dagger$}&\cellcolor{red!10}.28&\cellcolor{red!10}.32&\cellcolor{red!10}\textbf{.25$^\dagger$}&\cellcolor{red!10}\textbf{.17$^\dagger$}&\cellcolor{red!10}\textbf{.28$^\dagger$}&\cellcolor{red!10}\textbf{.28$^\dagger$}&\cellcolor{steel!30}.53$^\dagger$\\ 
\textsc{Keras-DNN/DSR-O2}&\cellcolor{steel!30}.52$^\dagger$&\cellcolor{steel!30}.52$^\dagger$&\cellcolor{steel!30}.52$^\dagger$&\cellcolor{red!10}\textbf{.43$^\dagger$}&\cellcolor{red!10}.38&\cellcolor{red!10}.37&\cellcolor{red!10}.28&\cellcolor{steel!30}.52$^\dagger$&\cellcolor{steel!30}.52$^\dagger$&\cellcolor{red!10}.47$^\dagger$&\cellcolor{red!10}.47$^\dagger$&\cellcolor{red!10}.40&\cellcolor{red!10}.37&\cellcolor{red!10}.45$^\dagger$&\cellcolor{steel!30}.52$^\dagger$&\cellcolor{steel!30}.52$^\dagger$&\cellcolor{red!10}\textbf{.43$^\dagger$}&\cellcolor{red!10}.45$^\dagger$&\cellcolor{red!10}.40&\cellcolor{red!10}.35&\cellcolor{red!10}.38\\ \hline
\textsc{Keras-DNN/Coffee-O1}&\cellcolor{red!10}.43&\cellcolor{steel!30}.53$^\dagger$&\cellcolor{steel!30}\textbf{.58$^\dagger$}&.50$^\dagger$&\cellcolor{red!10}.45&\cellcolor{steel!30}\textbf{.67$^\dagger$}&\cellcolor{steel!30}.55$^\dagger$&\cellcolor{red!10}.47$^\dagger$&.50$^\dagger$&\cellcolor{steel!30}\textbf{.68$^\dagger$}&\cellcolor{steel!30}.55$^\dagger$&\cellcolor{steel!30}\textbf{.57$^\dagger$}&\cellcolor{steel!30}\textbf{.60$^\dagger$}&\cellcolor{steel!30}\textbf{.58$^\dagger$}&\cellcolor{red!10}.38&\cellcolor{red!10}.43&\cellcolor{steel!30}.52$^\dagger$&\cellcolor{steel!30}.55$^\dagger$&\cellcolor{steel!30}\textbf{.58$^\dagger$}&\cellcolor{steel!30}\textbf{.67$^\dagger$}&\cellcolor{steel!30}.52$^\dagger$\\ 
\textsc{Keras-DNN/Coffee-O2}&\cellcolor{steel!30}\textbf{.58$^\dagger$}&\cellcolor{steel!30}\textbf{.58$^\dagger$}&\cellcolor{steel!30}\textbf{.58$^\dagger$}&\cellcolor{steel!30}\textbf{.58$^\dagger$}&\cellcolor{steel!30}\textbf{.57$^\dagger$}&\cellcolor{steel!30}\textbf{.57$^\dagger$}&\cellcolor{steel!30}\textbf{.57$^\dagger$}&\cellcolor{steel!30}\textbf{.58$^\dagger$}&\cellcolor{steel!30}\textbf{.57$^\dagger$}&\cellcolor{steel!30}\textbf{.58$^\dagger$}&\cellcolor{steel!30}\textbf{.57$^\dagger$}&\cellcolor{steel!30}\textbf{.57$^\dagger$}&\cellcolor{steel!30}.53$^\dagger$&.50$^\dagger$&\cellcolor{steel!30}\textbf{.58$^\dagger$}&\cellcolor{steel!30}\textbf{.58$^\dagger$}&\cellcolor{steel!30}.53$^\dagger$&\cellcolor{steel!30}\textbf{.57$^\dagger$}&\cellcolor{steel!30}.55$^\dagger$&\cellcolor{steel!30}\textbf{.58$^\dagger$}&\cellcolor{steel!30}\textbf{.57$^\dagger$}\\ \hline
\textsc{Keras-LSTM-O1}&\cellcolor{steel!30}\textbf{.68$^\dagger$}&\cellcolor{steel!30}\textbf{.58$^\dagger$}&\cellcolor{red!10}.47$^\dagger$&\cellcolor{red!10}.30&\cellcolor{red!10}.47$^\dagger$&\cellcolor{red!10}.47$^\dagger$&\cellcolor{red!10}.38&\cellcolor{red!10}.32&\cellcolor{steel!30}\textbf{.70$^\dagger$}&\cellcolor{red!10}.35&\cellcolor{red!10}.48$^\dagger$&\cellcolor{red!10}.32&\cellcolor{red!10}.42&\cellcolor{steel!30}\textbf{.58$^\dagger$}&\cellcolor{red!10}.47$^\dagger$&\cellcolor{red!10}.35&\cellcolor{steel!30}\textbf{.72$^\dagger$}&\cellcolor{red!10}.47$^\dagger$&\cellcolor{red!10}.32&\cellcolor{red!10}.45$^\dagger$&\cellcolor{red!10}.28\\ 
\textsc{Keras-LSTM-O2}&\cellcolor{red!10}.37&\cellcolor{red!10}.38&\cellcolor{red!10}.48$^\dagger$&\cellcolor{red!10}.45$^\dagger$&\cellcolor{red!10}.32&\cellcolor{red!10}.47$^\dagger$&\cellcolor{red!10}.38&\cellcolor{red!10}.42&\cellcolor{red!10}.30&\cellcolor{red!10}.33&\cellcolor{red!10}.42&\cellcolor{red!10}.28&\cellcolor{red!10}.48$^\dagger$&\cellcolor{red!10}.33&\cellcolor{red!10}.28&\cellcolor{steel!30}\textbf{.58$^\dagger$}&\cellcolor{red!10}.37&\cellcolor{red!10}.32&\cellcolor{red!10}.35&\cellcolor{red!10}.45$^\dagger$&\cellcolor{red!10}.35\\ 
\hline

$\#$ Best (over 16 cases)&4&2&1&2&2&2&3&3&3&2&0&1&4&3&6&2&1&1&1&3&2\\


\bottomrule
\end{tabular}
\centering
 \begin{tablenotes}
    \footnotesize
    \item $^\dagger$ denotes a statistically significant comparison with $p<0.05$. Other formats are the same as Table~\ref{tb:gain}.
    
    \end{tablenotes}
    \vspace{-0.3cm}
\end{table*}

%% file: related.tex
%

\section{Related Work}
\label{sec:related}


Broadly, optimizers for software configuration tuning can be classified into two categories: \textit{measurement-based} and \textit{model-based}.


\textbf{\textit{Measurement-based Optimizers:}} In measurement-based methods, the optimizer is directly used to measure the configuration on the software systems. Despite the expensiveness, the measurements can accurately reflect the good or bad of a configuration. The optimizer can be based on random search~\cite{DBLP:journals/jmlr/BergstraB12,DBLP:conf/sigmetrics/YeK03,DBLP:conf/sigsoft/OhBMS17}, hill climbing~\cite{DBLP:conf/www/XiLRXZ04,DBLP:conf/hpdc/LiZMTZBF14}, single-objective genetic algorithm~\cite{DBLP:conf/sc/BehzadLHBPAKS13,DBLP:conf/sigsoft/ShahbazianKBM20} and simulated annealing~\cite{DBLP:conf/icpads/DingLQ15,guo2010evaluating}, to name a few. Under such a single-objective model, various tricks have been applied. For example, some extend random search to consider a wider neighboring radius of the configuration structure, hence it is more likely to jump out from the local optima~\cite{DBLP:conf/sigsoft/OhBMS17}. Others rely on restarting from a different point, such as in restarted hill climbing, hence increasing the chance to find the ``right'' path from local optima to the global optimum~\cite{DBLP:conf/cloud/ZhuLGBMLSY17,DBLP:conf/www/XiLRXZ04}.

Our MMO model differs from all the above as it lies in a higher level of abstraction --- the optimization model --- as opposed to the level of optimization method. 


\textbf{\textit{Model-based Optimizers:}} Instead of solely using the measurements of software systems, the model-based methods apply a surrogate model (analytical~\cite{DBLP:conf/icse/Kumar0BB20,DBLP:conf/gecco/0001LY18,DBLP:journals/infsof/ChenLY19} or machine learning based~\cite{nair2018finding,DBLP:conf/mascots/JamshidiC16}) to evaluate configurations, which guides the search in an optimizer. The intention is to speed up the exploration of configurations as the model evaluation is rather cheap. Yet, it has been shown that the model accuracy and the availability of initial data can become an issue~\cite{DBLP:conf/cloud/ZhuLGBMLSY17}. Among others, Jamshidi and Casale~\cite{DBLP:conf/mascots/JamshidiC16} use Bayesian optimization to tune software configuration, wherein the search is guided by the Gaussian process regression trained from the data collected. Nair et al.~\cite{nair2018finding} follow a similar idea but a regression tree model is used instead.

Since MMO lies in the level of optimization model, it is complementary to the model-based methods in which the MMO would take the surrogate values as inputs instead of the real measurements.



\textbf{\textit{General Parameter Tuning:}} Optimizers proposed for the parameter tuning of general algorithms can also be relevant~\cite{DBLP:journals/ec/BlotMJH19,DBLP:journals/tec/HuangLY20,DBLP:series/sci/Bezerra0S20,DBLP:conf/ppsn/PushakH18}, including IRace~\cite{lopez2016irace}, ParamILS~\cite{DBLP:journals/jair/HutterHLS09}, SMAC~\cite{DBLP:conf/lion/HutterHL11}, GGA$++$~\cite{DBLP:conf/ijcai/AnsoteguiMSST15}, as well as their multi-objective variants, such as MO-ParamILS~\cite{DBLP:conf/lion/BlotHJKT16} and SPRINT-Race~\cite{DBLP:conf/gecco/ZhangGA15}. To examine a few examples, ParamILS~\cite{DBLP:journals/jair/HutterHLS09} relies on iterative local search --- a search procedure that may jump out of local optima using strategies similar to that of SA and SHC-r. Further, a key contribution is the capping strategy, which helps to reduce the need to measure an algorithm under some problem instances, hence saving computational resources. This is one of the goals that we seek to achieve too. Similar to Nair et al.~\cite{nair2018finding}, SMAC~\cite{DBLP:conf/lion/HutterHL11} uses Bayesian optimization but relying on a Random Forest model, which additionally considers the performance of an algorithms over a set of instances.

However, their work differs from ours in two aspects. 
	Firstly, general algorithm configuration requires to work on a set of problem instances, 
	each coming with different features. 
	The software configuration tuning, in contrast, is often concerned with tuning software system under a given benchmark (i.e., one instance)~\cite{nair2018finding,DBLP:conf/mascots/JamshidiC16,DBLP:conf/cloud/ZhuLGBMLSY17,Chen2018FEMOSAA}. Therefore, most of their designs for saving resources (such as the capping in ParamILS) were proposed to reduce the number of instances measured. Of course, it is possible to generalize the problem to consider multiple benchmarks as the same time, yet this is outside the scope of this paper. 
	Secondly, none of them works on the level of optimization model, and therefore our MMO is still complementary to their optimizers.


\textbf{\textit{Multi-Objectivization in SBSE:}} Multi-objectivization, which is the notion behind our MMO model, has been applied in other SBSE problems~\cite{DBLP:journals/tse/YuanB20,derakhshanfar2020good,DBLP:conf/ssbse/MkaouerKBC14,DBLP:conf/ssbse/SoltaniDPDZD18}. For example, to reproduce a crash based on the crash report, one can purposely design a new auxiliary objective, which measures how widely a test case covers the code, to be optimized alongside with the target crash distance~\cite{derakhshanfar2020good}. A multi-objective optimizer, e.g., NSGA-II, is directly used thereafter. A similar case can be found also for the code refactoring problem~\cite{DBLP:conf/ssbse/MkaouerKBC14}. However, during the tuning process, such a model, i.e., PMO in this paper, could waste a significant amount of the resources in optimizing the auxiliary objective, 
which is of no interest. 
This is a particularly unwelcome issue for software configuration tuning where the measurement is expensive. As we have shown in Section~\ref{sec:result}, PMO performs even worse than the classic single-objective model in most of the cases.







%% file: conclusion.tex
\section{Conclusion and Future Work}
\label{sec:con}





This paper tackles the local optimum issue in software configuration tuning from a different perspective --- 
multi-objectivizing the single objective optimization scenario. 
We do this by proposing a meta multi-objective model (MMO), 
at the level of optimization model (external part), 
as opposed to existing work that focuses on developing an effective single-objective optimizer (internal part).
We compare MMO with four state-of-the-art single-objective optimizers 
and the plain multi-objectivization model over various scenarios. 
The results reveal that 
the MMO model:


\begin{itemize}

\item can generally be more effective in overcoming local optima;
\item and do so by consuming less resources in most cases;
\item can be sensitive to the weight, but there exist some commonly best values.
\end{itemize}

The idea of MMO is essentially to rotate the original space of target and auxiliary objectives 
hence that solutions with good target objective value and various auxiliary objective values incomparable. 
In this geometrical transformation, 
the weight parameter determines how far in terms of the auxiliary objective solutions are incomparable, 
relative to the target objective.
A comparison with methods of the similar idea 
(e.g., select solutions with good target objective and diverse auxiliary objective values) 
can be beneficial as it can help answer an underlying question --- 
can maintain the diversity of the auxiliary objective help optimization of the target one.
This is one of our subsequent studies.
Another direction of future work is to add more auxiliary objective.
In this regard,
how to do the transformation in the 3D space is the key.
On top of the above, 
an adaptive weight adjustment approach for the MMO model,
as suggested by the findings from \textbf{RQ3}, 
is certainly more desirable,
which is worth investigating in depth.
